\documentclass{aa}

\usepackage{graphicx, psfig, times, amsfonts}

\newcommand{\less}{\raisebox{-1.1mm}{$\stackrel{<}{\sim}$}}
\newcommand{\more}{\raisebox{-1.1mm}{$\stackrel{>}{\sim}$}}
\newcommand{\msol}{\mbox{M$_{\odot}$}}

\newcommand{\msolyr}{{M$_{\odot}$}\,yr$^{-1}$ }

\newcommand{\ks}{km s$^{-1}$}

\newcommand{\M}{{\sc 2mass}}
\newcommand{\OG}{{\sc ogle}}
\newcommand{\DE}{{\sc denis}}

\begin{document}

\title{Mira Variables in the OGLE Bulge fields
\thanks{Table~1 is available in electronic form at the CDS via
anonymous ftp to cdsarc.u-strasbg.fr (130.79.128.5) or via
http://cdsweb.u-strasbg.fr/cgi-bin/qcat?J/A+A/. 
Figure~1 and the Appendices are available in the on-line edition of A\&A.}  
}

\author{
M.A.T. Groenewegen
\inst{1}
\and
J.A.D.L. Blommaert 
\inst{1,2}
}

\institute{
Instituut voor Sterrenkunde, K.U. Leuven, Celestijnenlaan 200B, 
B--3001 Leuven, Belgium 
\and
Sterrenkundig Observatorium, Universiteit Gent, Krijgslaan 281-S9, B--9000 Gent, 
Belgium }

\date{received: 2005,  accepted: 14 June 2005}

\offprints{Martin Groenewegen (groen@ster.kuleuven.ac.be)}

\authorrunning{Groenewegen \& Blommaert}
\titlerunning{Mira variables in the Bulge}

\abstract{
The 222~000 $I$-band light curves of variable stars detected by the
\OG\--{\sc ii} survey in the direction of the Galactic Bulge have been
fitted and have also been correlated with the \DE\ and \M\ {\it
all-sky release} databases and with lists of known
objects. Lightcurves, and the results of the lightcurve fitting
(periods and amplitudes) and \DE\ and \M\ data are presented for 2691
objects with $I$-band semi-amplitude larger than 0.45 magnitude,
corresponding to classical Mira variables. 
The Mira period distribution of 6 fields at similar longitude but
spanning latitudes from $-1.2$ to $-5.8$ are statistically
indistinguisable indicating similar populations with initial masses of 
1.5-2~\msol\ (corresponding to ages of 1-3~Gyr). A field at similar
longitude at $b = -0.05$ from Glass et al. (2001) does show a
significantly different period distribution, indicating the presence
of a younger population of 2.5-3 \msol\ and ages below $1$~Gyr.
The $K$-band period-luminosity relation is presented for the whole
sample, and for sub-fields. The zero point depends on Galactic
longitude. Simulations are carried out to show that the observed
dependence of the zero point with $l$, and the number of stars per
field are naturally explained using the model of disk and bulge stars
of Binney et al. (1997), for a viewing angle (major-axis Bar - axis
perpendicular to the line-of-sight to the Galactic Centre) of 
43 $\pm$ 17 degrees. The simulations also show that biases in the
observed zero point are small, $<0.02$ mag.
A comparison is made with similar objects in the Magellanic Clouds,
studied in a previous paper. The slope of the $PL$-relation in the
Bulge and the MCs agree within the errorbars. Assuming the zero point
does not depend on metallicity, a distance modulus difference of 3.72
between Bulge and LMC is derived. This implies a LMC DM of 18.21 for
an assumed distance to the Galactic Centre (GC) of 7.9 kpc, or,
assuming a LMC DM of 18.50, a distance to the GC of 9.0 kpc.
From the results in Groenewegen (2004) it is found for carbon-rich
Miras that the $PL$-relation implies a relative SMC-LMC DM of 0.38,
assuming no metallicity dependence. This is somewhat smaller than the
often quoted value near 0.50. Following theoretical work by Wood
(1990) a metallicity term of the form $M_{\rm K} \sim \beta \log Z$ is
introduced. If a relative SMC-LMC DM of 0.50 is imposed, $\beta = 0.4$
is required, and for that value the distance to the GC becomes 8.6
$\pm$ 0.7 kpc (for a LMC DM of 18.50), within the errorbar of the
geometric determination of 7.9 $\pm$ 0.4 kpc (Eisenhauer et al. 2003).
An independent estimate using the absolute calibration of Feast
(2004b) leads to a distance estimate to the GC of 8.8 $\pm$ 0.4 kpc.
}


\maketitle

\section{Introduction}

In the course of the micro lensing surveys in the 1990's the
monitoring of the Small and Large Magellanic Clouds has revealed an
amazing number and variety of variable stars. A big impact was felt
and is being felt in any area of variable star research, like Cepheids
and RR Lyrae stars. Also in the area of variability in red variables
(RVs) and AGB stars there has been remarkable progress. Wood et
al. (1999) and Wood (2000) were the first to identify and label
different sequences ``ABC'' thought to represent the classical Mira
sequence (``C'') and overtone pulsators (``A,B''), and sequence ``D''
which is not yet satisfactorily explained (Olivier \& Wood 2003, Wood
et al. 2004. Stars on these sequence are referred to as having Long
Secondary Periods--LSPs).  This view has subsequently been confirmed
and expanded upon by Noda et al. (2002), Lebzelter et al. (2002),
Cioni et al. (2003), Ita et al. (2004) and Kiss \& Bedding (2003, 2004),
Fraser et al. (2005). These works differ in the source of the
variability data ({\sc macho}, \OG, {\sc eros}, {\sc moa}), area (SMC
or LMC), associated infrared data (Siding Spring 2.3m, \DE, \M, {\sc
sirius}), and selection on pulsation amplitude or infrared colours. In
a recent paper, Groenewegen (2004; hereafter G04) analysed the \OG\
data in the SMC and LMC, and correlated these sources with the \DE\
and \M\ surveys. The paper discussed the variability properties of
three samples: about 2300 spectroscopically confirmed AGB stars,
around 400 previously known LPV variables and about 570 candidate 
dust-obscured AGB stars.

The present paper is an extension of the analysis in G04 to the \OG\
data in the direction of the Galactic Bulge (GB). Also for this area of
the sky, several papers exist that use the results of the micro 
lensing surveys and have extended previous classical works on Bulge 
variable stars, like those of Lloyd Evans (1976),
Glass \& Feast (1982), Whitelock et al. (1991), Glass et al. (1995;
hereafter GWCF), Alard et al. (1996) and Glass et al. (2001).

Alard et al. (2001; herafter ABC01) correlated ISOGAL sources within
the NGC 6522 and Sgr {\sc I} Baade windows with the MACHO database and
present a list of 332 stars with complete 4-band $V,R$ and [7],[15]
magnitudes. Schultheis \& Glass (2001) extended Alard et al. by also
considering the \DE\ and \M\ data in those fields in general, and for
the variables in particular. Glass \& Schultheis (2002; hereafter
GS02) investigated a sample of 174 M-giants in the NGC 6522 Baade
window and correlated them with \DE\, ISOGAL and MACHO. Many stars of
spectral type M5 and all M6 and later show variation, whereas subtypes
M1-M4 do not (see also Glass et al. 1999).

Glass \& Schultheis (2003; hereafter GS03) investigated the variable
stars in the NGC 6522 Baade's window using MACHO data, and also used
\DE\ IR data. Of the 1661 selected stars 1085 were found to be
variable. They present $K$-band $PL$-relations for sequences ``ABCD''.

Wray et al. (2004) investigated small amplitude red giants variables
in a sub-set of 33 \OG\ fields. They identified two groups that seem
to correspond to groups ``A-'' and ``B-'' in Ita et al. (2004; also
see G04).

In our paper we describe results on Mira variables selected from \OG\
Bulge fields. The paper is structured as follows. In Section~2 the
\OG, \M\ and \DE\ surveys are described. In Section~3 the model for
the lightcurve analysis is briefly presented. In the remaining of the
paper different results are described. The Period-Luminosity diagram
is discussed in Section~5. A description of the Mira population in
respect to the overall bulge population is given in Section 8. In
Section~9 we show that the Miras are distributed in a bar-like
structure and give the orientation. In the final section we give the
distance to the Galactic Centre, based on the period-luminosity
relation.

\section{The data sets}

The \OG\--{\sc ii} micro lensing experiment observed fourty-nine fields
in the direction of the GB. Each field has a size
14.2\arcmin$\times$57\arcmin\ and was observed in $BVI$, with an
absolute photometric accuracy of 0.01-0.02 mag (Udalski et al. 2002).
Table~\ref{tab-ogle} lists the galactic coordinates of the field
centers and the total number of sources detected in these fields.

Wozniak et al. (2002) present a catalog of about 222~000 variable
objects based on the \OG\ observations covering 1997-1999, applying
the difference image analysis (DIA) technique on the $I$-band data.
The data files containing the $I$-band data of the candidate variable
stars was downloaded from the \OG\ homepage
(http://sirius.astrouw.edu.pl/$^{\sim}$ogle/). According to 
Wozniak et al., the level of
contamination by spurious detections is about 10\%, but we presume
this level is much less at the brighter magnitudes of the LPVs
considered here.
Table~\ref{tab-ogle} lists the
number of detected variable stars per field (Wozniak et al. 2002).

The \DE\ survey is a survey of the southern hemisphere in $IJK_{\rm s}$ 
(Epchtein et al. 1999). The second data release available through
ViZier was used (The DENIS consortium, 2003). The 221801 \OG\ objects
were correlated on position using a 3\arcsec\ search radius and 59894
matches were found.

The \M\ survey is an all-sky survey in the $JHK_{\rm s}$ near-infrared
bands.  On March 25, 2003 the \M\ team released the all-sky point
source catalog (Cutri et al. 2003). The easiest way to check if a star
is included in the \M\ database is by uplinking a source table with
coordinates to the \M\ homepage. Such a table was prepared for the
221801 \OG\ objects and correlated on position using a 3\arcsec\
radius. Data on 182361 objects were returned.

\begin{table*}
\caption[]{
First entries in the electronically available table, which lists:
OGLE-field and number, the three fitted periods with errors and
amplitude (0.00 means no fit), mean $I_{\rm ogle}$, and associated \DE\
$IJK$ photometry with errors, and associated \M\ $JHK$ photometry with
errors (99.9 and 9.99 means no association, or no value).
}
\includegraphics[angle=+90, width=207mm]{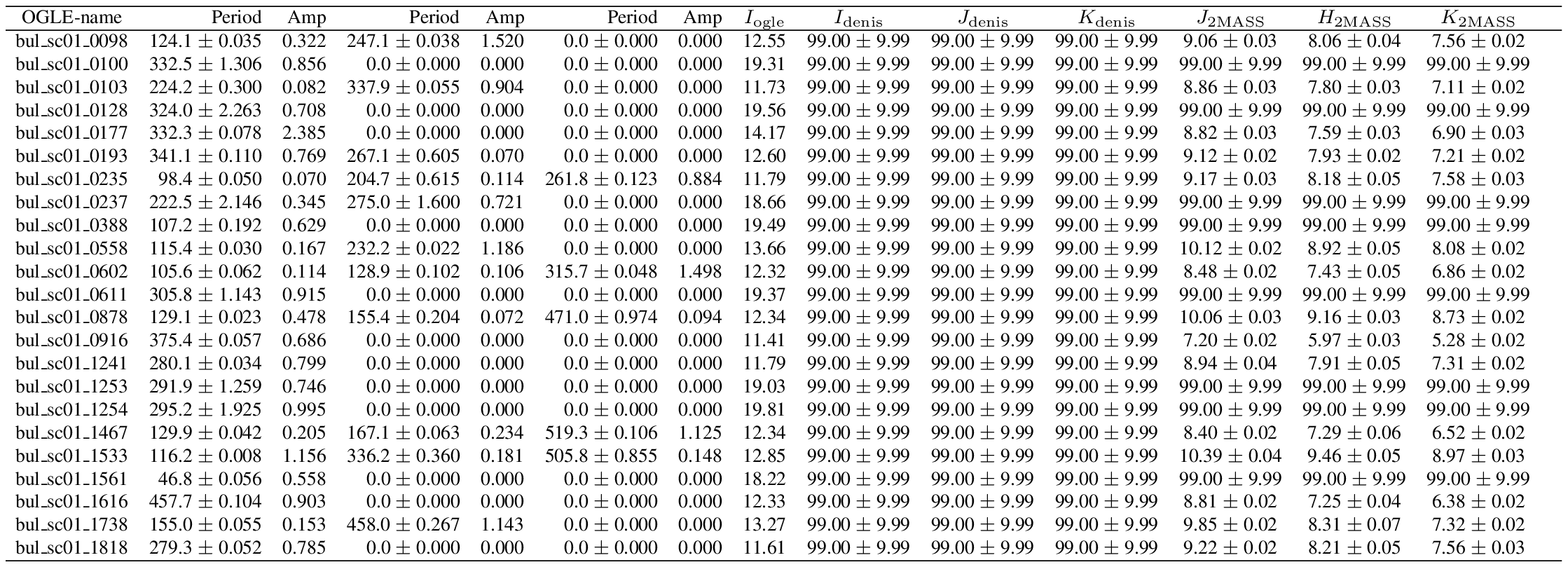}

\label{TAB-A}
\end{table*}

\begin{table}
\caption[]{
First entries in the electronically available table, which list:
OGLE-field and number, other names, spectral type and references.
ISOGAL sources from the official catalog (OGA03) have the prefix
``ISOGAL''.  ISOGAL sources studied in OOS03 have the prefix
``OOS03''. References in column~4 are given in the bibliography of the
main text.
}
\includegraphics[angle=+90, width=220mm]{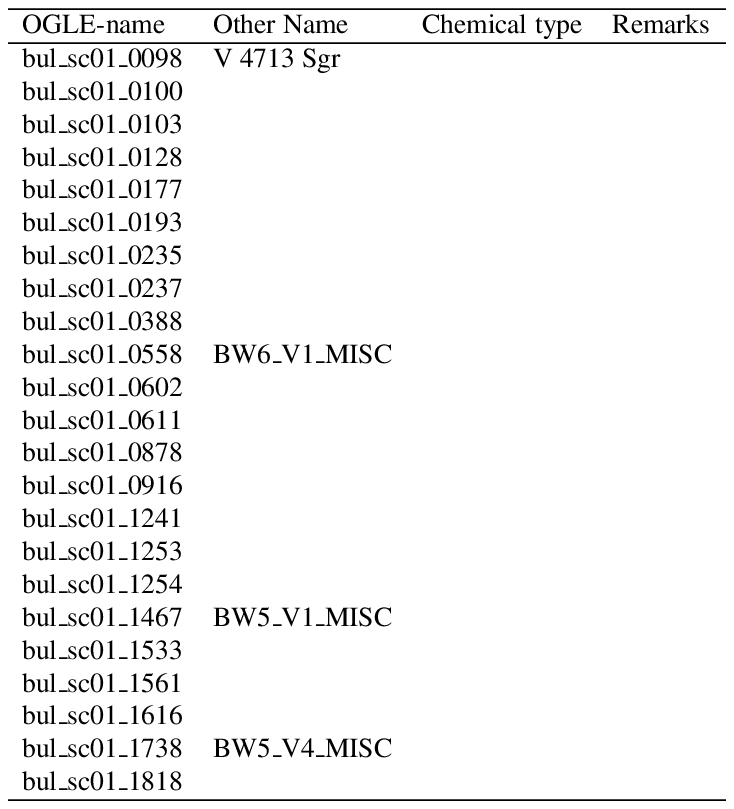}

\label{TAB-B}
\end{table}

\section{Lightcurve analysis}

The model to analyse the lightcurves is described in detail in
Appendices~A-C in G04.

Briefly, a first code (see for details Appendix~A in G04) sequentially
reads in the $I$-band data for the 222~000 objects, determines periods
through Fourier analysis, fits sine and cosine functions to the light
curve through linear least-squares fitting and makes the final
correlation with the pre-prepared \DE\ and \M\ source lists. All the
relevant output quantities are written to file.

This file is read in by the second code (see for details Appendix~B in G04). 
A further selection may be applied (typically on period, amplitude and
mean $I$-magnitude), multiple entries are filtered out (i.e. objects
that appear in different \OG\ fields), and a correlation is made with
pre-prepared lists of known non-LPVs and known LPVs or AGB stars. The
output of the second code is a list with LPV candidates.

The third step (for details see Appendix~C in G04) consists of a
visual inspection of the fits to the light curves of the candidate
LPVs and a literature study through a correlation with the {\sc simbad} 
database. Non-LPVs are removed, and sometimes the fitting is redone.  
The final list of LPV candidates is compiled.

Details on the small changes in the codes w.r.t. the implementation in
G04 are given in Appendix~A of the present paper.

\begin{figure*}
\includegraphics[width=175mm]{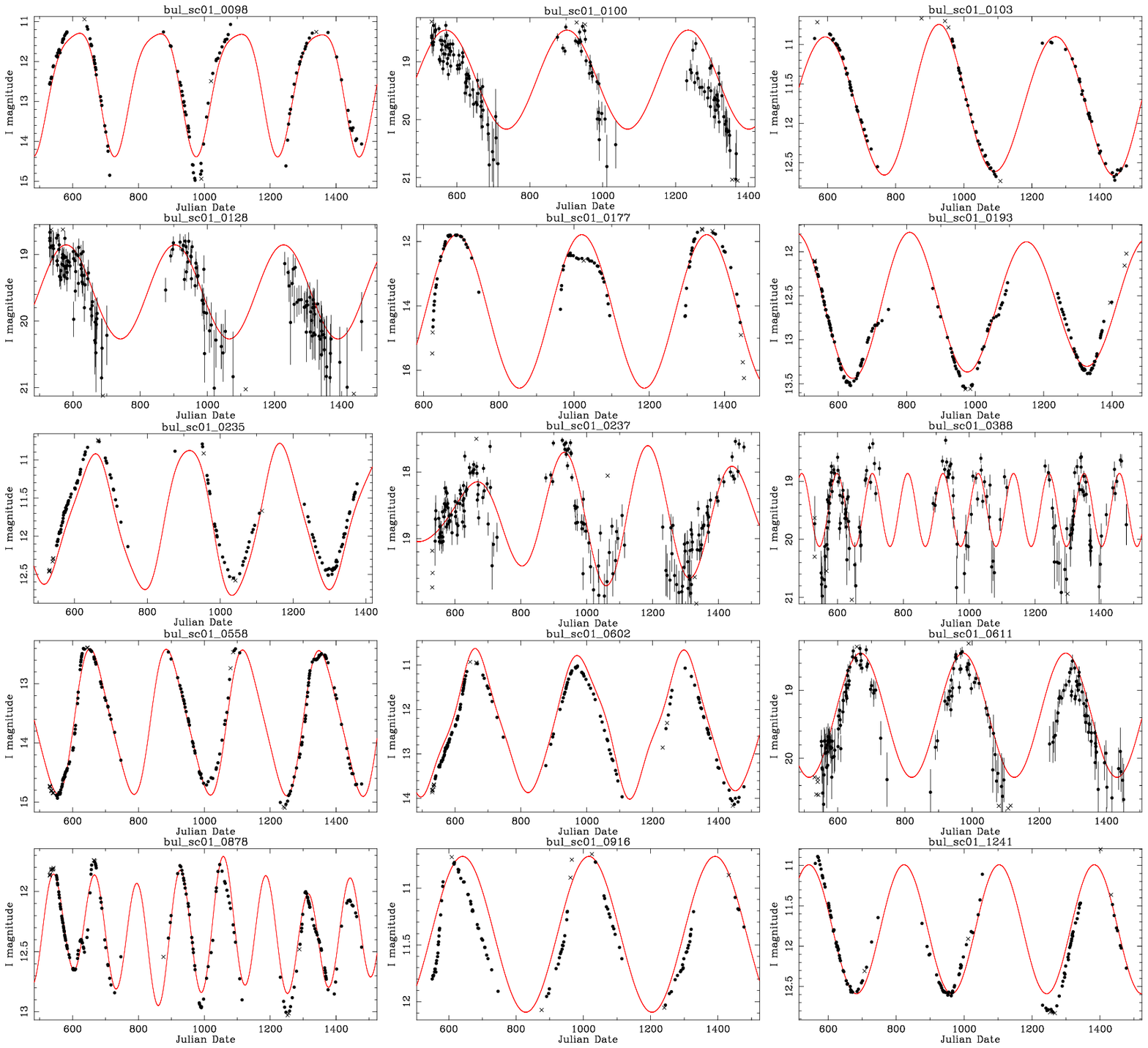}
\caption[]{
First entries of electronically available figure with all
lightcurves. The fit is indicated by the (red) solid line. Crosses
indicate data points not included in the fit.
}
\label{Fig-LC}
\end{figure*}

\section{Comparison of the datasets}

\subsection{Astrometry}

As in G04, the correlation between the \OG\ objects and known LPVs and
AGB stars, and known non-LPVs, is actually done in 2 steps. In the
first step the correlation is made (for a 3\arcsec\ search radius),
and the differences and spread in $\Delta$RA $\cos (\delta)$ and
$\Delta \delta$ are determined. These mean offsets are then applied in
most cases to make the final cross-correlation, and this usually
increases the number of matches. The results are listed in Table~\ref{tab-astro}.

\begin{table*}
\caption{Comparison of coordinates, and number of positional matches, before and after a correction was applied.}
\setlength{\tabcolsep}{1.0mm}
\begin{tabular}{lrrrrrrl} \hline
\small
OGLE  & $\Delta$RA $\cos (\delta)$ & $\Delta \delta$ & $N$ & $\Delta$RA $\cos (\delta)$ & $\Delta \delta$ & $N$ & remark \\ 
\hspace{1mm} -({\it other}) & & & & & & & \\ \hline 
IRAS       &  $ 0.14 \pm$ 0.94  &  $ 0.50 \pm$  1.44 &   80 &                  &                   &      & IRAS sources, not corrected \\
OGA03      &  $-0.80 \pm$ 0.85  &  $-0.16 \pm$  0.89 & 1309 & $-0.11 \pm$ 0.82 & $-0.02 \pm$  0.90 & 1319 & ISOGAL sources from Omont et al. (2003; OGA03) \\
OOS03      &  $-0.80 \pm$ 0.88  &  $-0.30 \pm$  0.95 &  800 & $-0.10 \pm$ 0.86 & $-0.03 \pm$  0.99 &  817 & ISOGAL sources from Ojha et al. (2003; OOS03) \\
GS02       &  $-0.02 \pm$ 0.58  &  $ 0.30 \pm$  0.61 &   79 & $-0.04 \pm$ 0.67 & $ 0.00 \pm$  0.61 &   80 & MACHO sources from Glass \& Schultheis (2002; GS02) \\
ABC01      &  $-0.82 \pm$ 0.98  &  $-0.70 \pm$  0.96 &  293 & $-0.17 \pm$ 1.08 & $-0.12 \pm$  1.01 &  320 & ISOGAL sources from Alard et al. (2001; ABC01) \\
OGLE-I     &  $-0.08 \pm$ 0.70  &  $-0.18 \pm$  0.44 &  728 &                  &                   &      & \OG\--{\sc i} sources, not corrected \\
BMB        &  $-0.04 \pm$ 0.70  &  $-0.52 \pm$  0.90 &  282 & $+0.10 \pm$ 0.79 & $-0.08 \pm$  0.91 &  286 & sources from Blanco et al. (1984; BMB) \\
TLE        &  $-0.08 \pm$ 0.84  &  $-0.25 \pm$  1.43 &   21 &                  &                   &      & sources from  Lloyd-Evans (1976), not corrected \\
B84        &  $-0.46 \pm$ 0.75  &  $ 0.50 \pm$  0.97 &   61 & $-0.02 \pm$ 0.76 & $ 0.08 \pm$  1.05 &   63 & sources from Blanco (1984; B84) \\
\hline
\end{tabular}
\label{tab-astro}
\end{table*}

\subsection{Photometry}

As in G04, a comparison was made between the (mean) \OG\ $I$ and the
(single-epoch) \DE\ $I$, and between the (single-epoch) \DE\ $JK$ and
the (single-epoch) \M\ $JK$ magnitudes. This was done by selecting
those objects with an amplitude in the $I$-band of $<0.05$ mag.

\begin{figure}
\centerline{\psfig{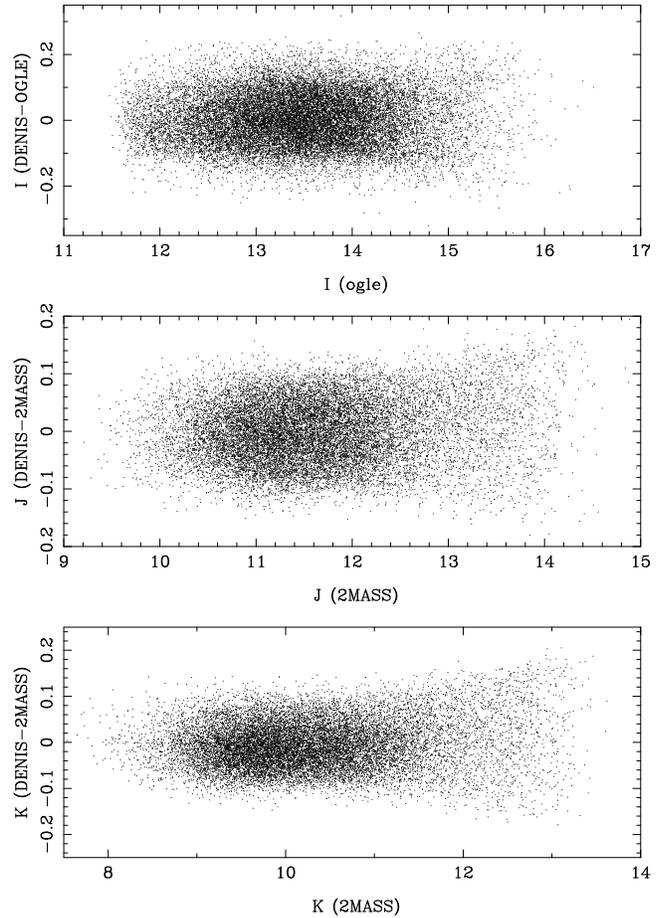}}
\caption[]{
Difference in photometry, {\em after} the following offsets have been
applied: $I$(denis-ogle) = $-0.03$, $J$(denis-2mass)= $-0.02$, $K$(denis-2mass)= $-0.03$.
}
\label{Fig-PHT}
\end{figure}

Figure~\ref{Fig-PHT} shows the final results when offsets
$I$(denis-ogle) = $-0.03$, $J$(denis-2mass)= $-0.02$, and
$K$(denis-2mass)= $-0.03$ are applied. The latter values are
consistent with those derived in OOS03 based on the \M\ second
incremental data release who found $J$(denis-2mass)= $-0.02 \pm 0.09$,
and $K$(denis-2mass)= $-0.00 \pm 0.07$.

\section{Period-Luminosity relations}

The full machinery outlined in Section~3 was performed.  As in G04,
all derived periods are given in Table~\ref{TAB-A} and are shown in
Figure~\ref{Fig-LC}. The present paper discusses only objects which
have at least one period with an $I$-band amplitude larger than 0.45
magnitudes\footnote{The amplitude, $A$, is used in the
mathematical sense in the present paper, $y= A \; \sin x$. 
The peak-to-peak amplitude is 0.90 mag.}, i.e. classical Mira variables
(e.g. Hughes 1989). After visual inspection of the lightcurves a
sample of 2691 such objects remain. The number of objects per field is
listed in the last column of Tab.~\ref{tab-ogle}.

Table~\ref{TAB-A} lists the stars in the sample, the fitted periods
with errors and amplitudes, and the \DE\ and \M\ photometry of the
associated sources. Table~\ref{TAB-B} lists alternative names and
references from the literature.
Figure~\ref{Fig-LC} presents lightcurves and their fits.

In the discussion that follows, magnitudes are de-reddened using the
$A_{\rm V}$ values that correspond to the respective \OG\ field taken
from Sumi (2004; and $A_{\rm V}$ = 6.0 for the field 44 that they do
not discuss), and selective reddenings of $A_{\rm I}/A_{\rm V} =
0.49$, $A_{\rm J}/A_{\rm V} = 0.27$, $A_{\rm H}/A_{\rm V} = 0.20$,
$A_{\rm K}/A_{\rm V} = 0.12$ (Draine 2003) and implicitly assuming
that all objects suffer this reddening value (i.e. ignoring
differential reddening within a field, and ignoring that foreground
and background objects would suffer a different reddening). Sumi's
method is based on the absolute magnitude of the Red Clump giants and
is absolute calibrated using the $(V-K)$ colours of 20 RR Lyrae stars
in Baade's window. Popowski et al. (2003) present an extinction map
(over 9000 resolution elements of 4x4 arc minute size) towards the GB
based on MACHO $V,R$ photometry, under the assumption that
colour-magnitude diagrams would look similar in the absence of
extinction. For the centre of the \OG\ fields it was checked if there
was a tile in the Popowski et al. set within 0.05 degrees distance.
For those, the value of the visual extinction has been listed, next to
the value in Sumi in Table~\ref{tab-ogle}. The rms difference $A_{\rm V}$ 
for the 21 field with values from both references is 0.18.
Finally, Schultheis et al. (1999b) presented a reddening map for the
inner GB comparing \DE\ $J,K$ photometry to isochrones. Table~\ref{tab-ogle} 
lists for two \OG\ fields the values they find: SC44 which was not
considered by Sumi (2004), and SC5 for which Sumi derive a larger
$A_{\rm V}$ than Schultheis et al.: 5.73 versus 4.13. 

In the further discussion we only use periods that
fulfil the following conditions are used in the calculations (with
$\Delta P$ the error in the period): $\Delta P/P < 0.01$ for $P<
500^d$; $\Delta P < 5^d$ for $500^d < P < 800^d$ and $\Delta P <
1.5^d$ for $P > 800^d$. The latter constraint is necessary because the 
long periods become comparable to the length of the dataset.

\begin{figure*}
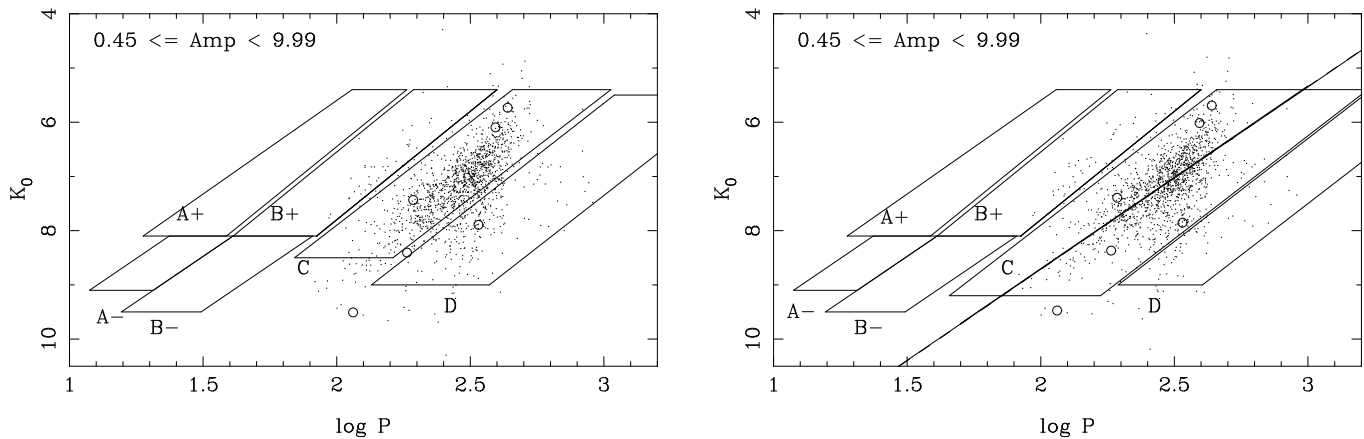


\begin{minipage}{0.48\textwidth}
\resizebox{\hsize}{!}{\includegraphics{K-P_ShiftedBy4Mag.ps}}
\end{minipage}
\hfill
\begin{minipage}{0.48\textwidth}
\resizebox{\hsize}{!}{\includegraphics{K-P_BiggerC.ps}}
\end{minipage}

\caption[]{
$K$-band $PL$-relation for periods with an $I$-band amplitude larger
than 0.45 mag and $(J-K)_0 < 2.0$. Left panel: Sequences/Boxes ``A+,
A-, B+, B-, C, D'' taken from G04 for the LMC and shifted by -4.0 in
distance modulus.  Right panel: enlarged 
Box ``C'' (and contracted Box ``D'',) to define the
region for which the $PL$-relations will be computed. 
Only periods from Table~\ref{TAB-A} that fulfil $\Delta P/P < 0.01$ for
$P< 500^d$; $\Delta P < 5^d$ for $500^d < P < 800^d$ and $\Delta P <
1.5^d$ for $P > 800^d$ are plotted and used in the analysis.
known M-stars are indicated by open circels.
The line in the right-hand panel indicates the $PL$-relation of Equation~1.
}
\label{Fig-PL}
\end{figure*}

Figure~\ref{Fig-PL} shows the $K$-band $PL$-relation for all periods
which have an $I$-band amplitude larger than 0.45 magnitude and
$(J-K)_0 < 2.0$ among the 2691 stars. The cut in $(J-K)$ colour is
needed to prevent that the $K$-magnitude is affected by circumstellar
extinction, as shown in G04. Like G04, the $K$-magnitude is on the \M\
system, and is the average of the \DE\ and \M\ photometry. In
particular, if both \DE\ and \M\ $K$-band data is available, the \DE\
data point is corrected as explained above (i.e. 0.03 mag added), and
averaged with the \M\ data point. This should take out some of the
scatter in the $PL$-diagram, as the effect of the variability in the
$K$-band is reduced. If only \DE\ is available, the corrected value is
used.  In the left-hand panel the boundaries of the boxes ``A-, A+,
B-, B+, C, D'' have been taken from G04, but shifted by $-4.0$ to
account for the approximate difference in distance modulus (DM), as,
e.g., follows from the recent determination of 7.9 $\pm$ 0.4 kpc
(Eisenhauer et al. 2003) for the distance to the GC, and 18.50 for the
DM to the LMC (e.g. recent reviews by Walker 2003, Feast 2004a).

There is a reasonably well defined sequence in Box ``C'', but when
compared to the similar figure for the SMC and LMC in G04 (his
figure~3) some differences can be remarked as well. In particular, for
the present Bulge sample there are a few objects located in Box
``B+'', and in particular many in Box ``D''. In the SMC and LMC, for
this cut in amplitude, there are none in Box ``B+'' and few in
``D''. Several issues may play a role. Applying a certain cut in
amplitude may sample slightly different variables in SMC, LMC and
Bulge.  Figure~3 in G04 clearly shows how lowering the cut in
amplitude results in a populating Box ``B+'' and then ``A+'', and
increases the number of objects in ``D''. Another effect is the
possible contribution of objects in the foreground and background of
the Bulge, the depth of the Bulge, and fourthly, the orientation of
the Bar, as the \OG\ fields span 20 degrees in longitude (this last
effect will be discussed later). Finally, the difference in DM may be
different from the adopted value of 4.0.

To verify if the objects in Box ``D'' actually show LSP, they were all
visually inspected, and in fact few have, in agreement with the
finding for LMC and SMC (for amplitudes $>$0.45 mag). This would call
for a enlargement of Box ``C'' to properly sample the $PL$-relation of
the large amplitude (Mira) variables. To define this enlarged box the
$PL$-relation was inspected for each field independently.  The
right-hand panel in Figure~\ref{Fig-PL} shows the finally adopted
boundaries of Box ``C'', which implies that Box ``D'' has contracted. 
Stars inside this redefined Box will be used
to define the $PL$-relation. The $K$-band $PL$-relation
is determined to be: 
\begin{equation}
m_{\rm K}= (-3.37 \pm 0.09)\; \log P + (15.44 \pm 0.21)
\end{equation}
with an rms of 0.42 and based on 1292 stars, and is shown in Fig.~\ref{Fig-PL}.
The value of the slope is consistent with the median value when the
$PL$-relation is determined for all fields individually.

For reference, fitting all stars in Figure~\ref{Fig-PL}, for a fixed
slope of $-3.37$ results in a ZP of 15.47 $\pm$ 0.55.

\begin{table*}
\caption{Properties of the \OG-fields}


\begin{tabular}{rrrrrlr} \hline
BUL\_SC & $l$ & $b$ & Total$^{(a)}$ & Variable$^{(b)}$ & $A_{\rm V}$$^{(c)}$ 
                                          &  LPV$^{(d)}$  \\ 
        &   &   &    &      &             &           \\ \hline

1  &  1.08 & -3.62 & 730   & 4597 & 1.68 / 1.49 &  42 \\
2  &  2.23 & -3.46 & 803   & 5279 & 1.55 / 1.65 &  48 \\
3  &  0.11 & -1.93 & 806   & 8393 & 2.89        & 115 \\
4  &  0.43 & -2.01 & 774   & 9096 & 2.59 / 2.94 &  86 \\
5  & -0.23 & -1.33 & 434   & 7257 & 5.73 / -- / 4.13       & 118 \\
6  & -0.25 & -5.70 & 514   & 3211 & 1.37        &  47 \\
7  & -0.14 & -5.91 & 463   & 1618 & 1.33 / 1.28 &  21 \\
8  & 10.48 & -3.78 & 402   & 2331 & 2.14        &   8 \\
9  & 10.59 & -3.98 & 330   & 1847 & 2.08        &  21 \\
10 &  9.64 & -3.44 & 458   & 2499 & 2.23        &  16 \\
11 &  9.74 & -3.64 & 426   & 2256 & 2.27        &  18 \\
12 &  7.80 & -3.37 & 535   & 3476 & 2.29 / 2.20 &  33 \\
13 &  7.91 & -3.58 & 570   & 3084 & 2.06 / 1.82 &  21 \\
14 &  5.23 &  2.81 & 619   & 4051 & 2.49        &  51 \\
15 &  5.38 &  2.63 & 601   & 3853 & 2.77        &  71 \\
16 &  5.10 & -3.29 & 700   & 4802 & 2.15 / 2.23 &  45 \\
17 &  5.28 & -3.45 & 687   & 4690 & 1.94 / 2.29 & 167 \\
18 &  3.97 & -3.14 & 749   & 5805 & 1.83        &  55 \\
19 &  4.08 & -3.35 & 732   & 5255 & 2.74        &  51 \\
20 &  1.68 & -2.47 & 785   & 5910 & 1.94 / 2.02 &  64 \\
21 &  1.80 & -2.66 & 883   & 7449 & 1.83 / 1.78 &  60 \\
22 & -0.26 & -2.95 & 715   & 5589 & 2.74        &  70 \\
23 & -0.50 & -3.36 & 723   & 4815 & 2.70        &  60 \\
24 & -2.44 & -3.36 & 612   & 4304 & 2.52        &  56 \\
25 & -2.32 & -3.56 & 622   & 3046 & 2.34        &  61 \\
26 & -4.90 & -3.37 & 728   & 4713 & 1.86        &  39 \\
27 & -4.92 & -3.65 & 691   & 3691 & 1.69        &  46 \\
28 & -6.76 & -4.43 & 406   & 1472 & 1.64        &  16 \\
29 & -6.64 & -4.62 & 492   & 2398 & 1.53        &  36 \\
30 &  1.94 & -2.84 & 762   & 6893 & 1.91 / 1.78 &  50 \\
31 &  2.23 & -2.94 & 790   & 4789 & 1.81 / 1.74 &  59 \\
32 &  2.34 & -3.14 & 797   & 5007 & 1.61 / 1.82 &  54 \\
33 &  2.35 & -3.66 & 739   & 4590 & 1.70 / 1.82 & 105 \\
34 &  1.35 & -2.40 & 961   & 7953 & 2.52 / 2.32 &  60 \\
35 &  3.05 & -3.00 & 771   & 5169 & 1.84 / 2.20 &  39 \\
36 &  3.16 & -3.20 & 873   & 8805 & 1.62 / 1.52 &  38 \\
37 &  0.00 & -1.74 & 664   & 8367 & 3.77        & 108 \\
38 &  0.97 & -3.42 & 710   & 5072 & 1.83 / 1.94 &  57 \\
39 &  0.53 & -2.21 & 784   & 7338 & 2.63 / 2.70 &  99 \\
40 & -2.99 & -3.14 & 631   & 4079 & 2.94        &  55 \\
41 & -2.78 & -3.27 & 603   & 4035 & 2.65        &  49 \\
42 &  4.48 & -3.38 & 601   & 4360 & 2.29        &  32 \\
43 &  0.37 &  2.95 & 474   & 3351 & 3.67        & 112 \\
44 & -0.43 & -1.19 & 319   & 7836 & 6.0 / -- / 6.00        & 132 \\
45 &  0.98 & -3.94 & 627   & 2262 & 1.64 / 1.53 &  32 \\
46 &  1.09 & -4.14 & 552   & 2057 & 1.71 / 1.65 &  26 \\
47 & -11.19 & -2.60 & 301  & 1152 & 2.60        &  12 \\
48 & -11.07 & -2.78 & 287  &  973 & 2.35        &  12 \\
49 & -11.36 & -3.25 & 251  &  826 & 2.09        &  18 \\
total &     &     & 30490  & 221701     &       & 2691       \\ 
\hline
\end{tabular}
\label{tab-ogle}

(a) Total number of objects detected in the field. From Udalski et
al. (2002), in units of $10^3$ objects

(b) Total number of candidate variable stars. From Wozniak et al. (2002).

(c) Visual extinction. From Sumi (2004), except for SC44,
where $A_V$ = 6.0 has been adopted based on the proximity to SC5.
The second value--when listed--comes from Popowski et al. (2003).
The third value--when listed--comes from Schultheis et al. (1999b).

(d) Total number of LPVs.

\end{table*}

\section{Historical versus current periods}

Table~\ref{Tab-PoldPnew} compares the period derived in the present
paper (the one with the largest amplitude) with values derived in the
literature. There are three cases where a previously determined period
may be a harmonic of the present period but overall there is good
agreement between periods. In the 12 cases where there is a period
available from LE76 (with the photographic plates taken between 1969
and 1971, hence 28 years of time difference with \OG) there is no
clear case for a star that changed period. By comparison, G04 found
that about 8\% of LMC variables changed their period by more than 10\%
over a about 17 year timespan. To find 0 out of 12 in the present
sample is consistent with this.

\begin{table*}  
\caption{Current Period compared to values listed in the literature }
\begin{tabular}{llcl} \hline
OGLE-name       & Periods  & Remark &  \\ \hline
bul\_sc01\_0558 & 232.2 (OGLE-II), 217.0, Mira (OGLE-I)   &  &  \\
bul\_sc01\_1467 & 519.3 (OGLE-II), 265.0, Mira (OGLE-I)   & harmonic ? &  \\  
bul\_sc01\_1738 & 458.0 (OGLE-II), 227.5, Mira (OGLE-I)   & harmonic ? &  \\  
bul\_sc01\_2079 & 524.7 (OGLE-II), 520.0, Mira (OGLE-I)   &  &  \\  
bul\_sc20\_0832 & 245.9 (OGLE-II), 254.1: (ABC01), 235 (LE76), 336 (GWCF) &  &  \\
bul\_sc20\_0975 & 167.7 (OGLE-II), 145.8 (ABC01) &  &  \\  
bul\_sc20\_1189 & 220.0 (OGLE-II), 121.3 (ABC01) & harmonic ? &  \\  
bul\_sc20\_1292 &  87.7 (OGLE-II),  88.1 (ABC01) &  &  \\  
bul\_sc20\_1761 & 231.6 (OGLE-II), 107.9 (ABC01), 240 (LE76), 235 (GWCF) &  & \\  
bul\_sc20\_1826 & 331.7 (OGLE-II), 330: (LE76) &  &  \\  
bul\_sc20\_1928 & 297.6 (OGLE-II), 300 (LE76), 293 (GWCF) &  &  \\
bul\_sc20\_2013 & 453.3 (OGLE-II), 474.2: (ABC01), 500:: (LE76), 480 (GWCF) &  &  \\
bul\_sc20\_2269 & 409.3 (OGLE-II), 400: (LE76), 383 (GWCF) &  &  \\  
bul\_sc20\_2291 & 311.2 (OGLE-II), 306.9: (ABC01), 315 (LE76), 308 (GWCF) &  &  \\ 
bul\_sc34\_3759 & 289.8 (OGLE-II), 293.8 (ABC01), 265 (LE76), 237 (GWCF) &  &  \\
bul\_sc45\_0704 & 405.3 (OGLE-II), 430 (LE76), 467, Mira (OGLE-I), 413 (GS03) &  &  \\
bul\_sc45\_1068 & 115.2 (OGLE-II), 115 (LE76), 116 (GCVS), 116 (GS03) &  &   \\  
bul\_sc45\_1586 & 193.3 (OGLE-II), 190 (LE76), 195 (GS03)  &  &   \\  
bul\_sc46\_0866 & 322.1 (OGLE-II), 305 (LE76), 320 (GS03) &  & \\  
bul\_sc46\_1163 & 343.7 (OGLE-II), 329.0, Mira (OGLE-I)  &  &  \\  
\hline
\end{tabular}
\label{Tab-PoldPnew}
\end{table*}

\section{Colour-colour diagrams}

\begin{figure*}
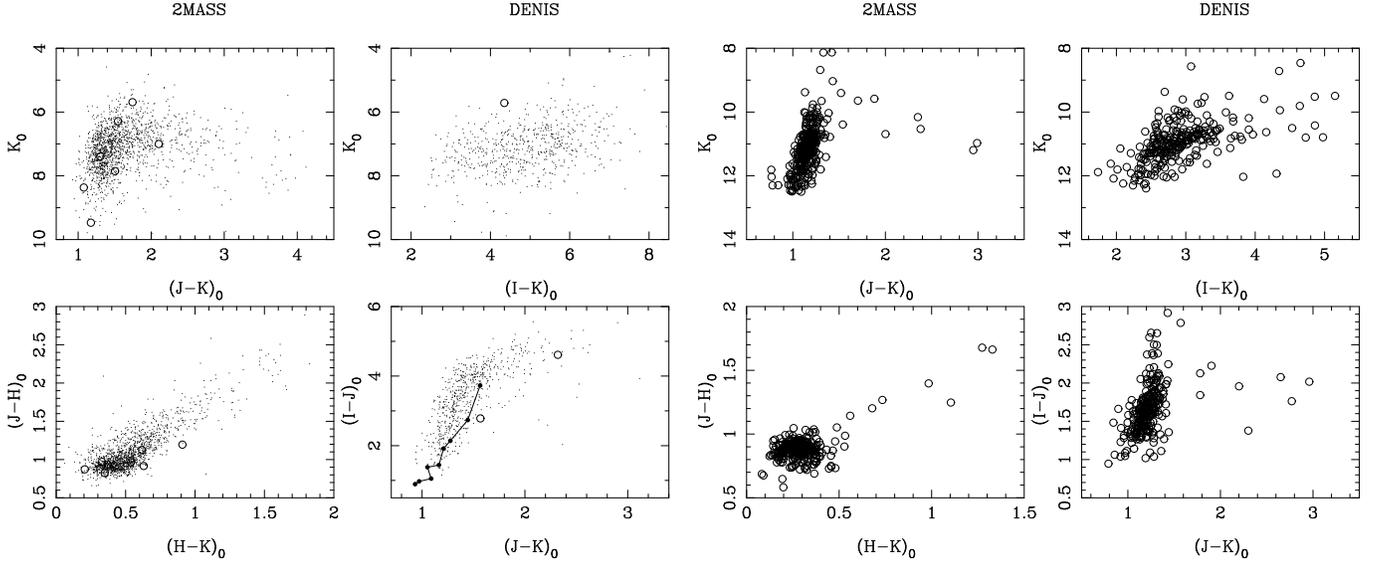


\begin{minipage}{0.49\textwidth}
\resizebox{\hsize}{!}{\includegraphics{CC_BUL_noProjCorr.ps}}
\end{minipage}
\hfill
\begin{minipage}{0.49\textwidth}
\resizebox{\hsize}{!}{\includegraphics{CC_LMC_Mstars.ps}}
\end{minipage}

\caption[]{
Colour-Magnitude and colour-colour diagrams using \M\ and \DE\
photometry for the Bulge stars (left), and spectroscopically confirmed
M-stars in the LMC (right, from G04).
The bullets connected by a line in the Bulge DENIS $(I-J) - (J-K)$
diagram are the average colours of M1, M2, .., M6, M6.5, M7, M8 giants
in the NGC 6522 Baade's window (Blanco 1986, Glass \& Schultheis 2002).
}
\label{Fig-CC}
\end{figure*}

The \M\ and \DE\ Colour-colour and colour-magnitude diagrams are shown
in Fig.~\ref{Fig-CC}, together with that of spectroscopically
confirmed M-stars in the LMC (see also Figure~12 in G04). There appear
to be more redder stars in the Bulge sample, but this is likely due to
a under representation in the LMC sample as this was restricted to
spectroscopically known M-stars (i.e. in general optically selected). 
The sample of candidate infrared-selected AGB stars in the LMC
[Figure~12 in G04] does cover the $(I-K)$ and $(J-K)$ colour range
observed in the Bulge).

The other main difference is that the Bulge stars are redder by $\sim 2$ 
mag in $(I-J)_0$ compared to both LMC and SMC stars, as was also shown 
by Lebzelter et al. (2002) in a comparison of LMC and Bulge variable stars.  
As the
diagrams involving $J, H, K$ colours appear similar, it seems that this
difference in $(I-J)$ must be due to a difference in $I$. The $I$-band 
measurements of M stars is strongly affected by the TiO and VO
molecular absorption features (Lan\c{c}on \& Wood 2000). It is expected that 
for larger metallicities these lines will be stronger (Schultheis et al. 
1999a) which will lead to redder $(I-J)$ colours.

The bullets connected by a line in the Bulge \DE\ $(I-J) - (J-K)$
colour-colour diagram are the average colours of M1, M2, .., M6, M6.5,
M7, M8 giants in the NGC 6522 Baade's window (Blanco 1986, GS02). 
There is a spread of typically 0.3-0.5 mag in $(I-J)$ and 0.2-0.3 mag
in $(J-K)$ around these means, and there is only 1 M8 giant in their
sample. The colours of the Miras follow those of normal giants well
until M6.5, when the Miras become redder in $(I-J)$.

There are also stars redder in $(J-K)$ than the single M8 star in the
sample of GS02, indicating either the presence of later spectral types
or the on-set of circumstellar reddening.

The conclusion of GS02 that ``Many M5 and all stars M6 and later show
variation, whereas subtypes (M1-M4) do not'' is confirmed here, as
there are essentially no objects located in the region of the \DE\
$(I-J) - (J-K)$ colour-colour diagram occupied by spectral types of M4
and earlier.

\section{Mira bulge population as function of latitude}

\begin{figure}
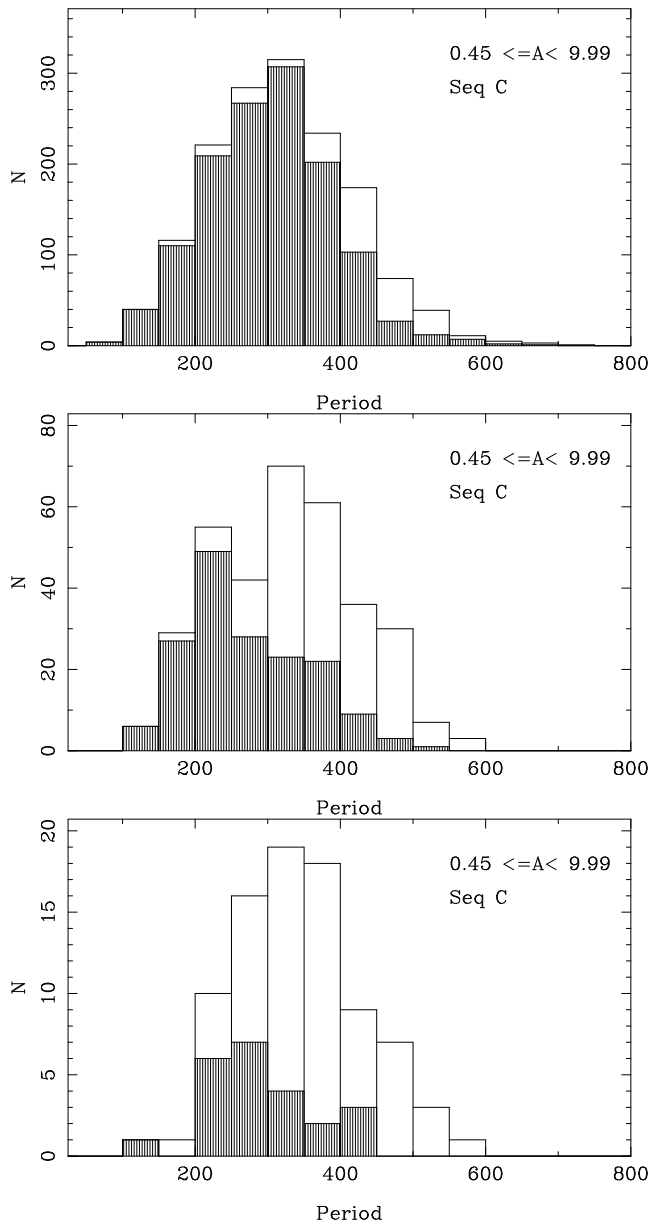

\centerline{\psfig{figure=PerDistbul_0.45_9.99.ps,width=8.5cm}}

\centerline{\psfig{figure=PerDistLmc_C_0.45_9.99_MIRA.ps,width=8.5cm}}

\centerline{\psfig{figure=PerDistSmc_C_0.45_9.99_MIRA.ps,width=8.5cm}}
\caption[]{
Period distribution of large amplitude variables in box ``C'', for
Bulge (top), LMC (middle), and SMC (bottom).  For the shaded histogram
only stars with $(J-K)_0 < 2.0$ have been included.  }
\label{Fig-PerD}
\end{figure}

Figure~\ref{Fig-PerD} shows the period distribution of Miras in Box ``C''. 
A distinction is made between all Miras and those with $(J-K)_0 <2.0$
(dashed histograms). The latter selection minimises any influence of
circumstellar extinction.  For comparison, the period distribution of
LMC and SMC Miras is also shown\footnote{Derived following the
implementation of the code and definition of the ``boxes'' as in G04,
and applying the same selection criteria as in the present paper,
i.e. $I$-amplitude larger than 0.45 mag.
}. 
The Kolmogorov-Smirnov (KS) test is performed to indicate that the
probability that the period distributions of Bulge-LMC, Bulge-SMC,
LMC-SMC are the same for all stars (those with $(J-K)_0<2$, respectively) 
is, respectively 0.36 (10$^{-8}$), 0.05 (0.31) and 0.05 (0.05).

Any difference, in particular between Bulge and LMC period
distribution, is difficult to quantify further as this depends in a
complicated way on the Star Formation History and evolutionary tracks
($T_{\rm eff}$ - Luminosity - Mass - metallicity).
Regarding the period distribution of Bulge Miras as such, previous
studies are limited to selected small fields (e.g. TLE, GWCF, Glass et
al. 2001). Whitelock et al. (1991) present the period distribution of
about 140 IRAS sources
but no direct comparison is possible because of the difference in
the selection criteria of the samples.

Figure~\ref{Fig-PerField} shows the period distribution of selected
fields with very similar longitudes that cover a range in latitudes
(the stars with $(J-K)_0 < 2.0$ are shown as dashed histogram again).
To add a field even closer to the GC than surveyed by \OG\ the data in
Glass et al. (2001, 2002) is considered on a field centered on $l =
-0.05, b = -0.05$. They present the results of a $K$-band survey of 24
$\times$ 24 arcmin$^2$ for LPVs down to $K \sim 12.0$. From the list
of 409 stars, 14 were removed because of double entries, quality index
of zero, uncertain or no listed period or amplitude. The coordinates
were uploaded to the IPAC webserver and \M\ data within 2.5\arcsec\
was retrieved for these 395 stars to get information on the $(J-K)$
colour.  As an additional check, and to eliminate multiple stars
within the search circle, it was verified that the single-epoch \M\
$K$ magnitude is consistent with the mean $K$-magnitude and amplitude
listed in Glass et al. For 345 stars \M\ data is available. The
magnitudes are corrected for interstellar reddening using the
extinction map of the inner GB at 2\arcmin\ resolution by Schultheis
et al. (1999b). The extinction value of the nearest available grid
point in this map is taken. The extinction values range between 18.5
and 30.4 with a mean of 24.7. The top panel in
Figure~\ref{Fig-PerField} list 333 stars with $K > 7$ (to eliminate 3
very likely foreground objects), and $K$-band amplitude larger than
0.35 (to correspond roughly to the cut in $I$-band amplitude of 0.45 mag),
and 88 (the histogram with slanted hatching), or 236 (dashed
histogram) stars which also have $(J-K)_0 < 2.0$.
The latter sample is the one that results when the reddening values
from Schultheis et al. are multiplied by 1.35.  They mention
themselves that the reddening may be underestimated in the direction
of the GC because of $J$-band non-detections. For the one field in
common, their value is a factor 1.3-1.4 smaller than derived by Sumi
(2004). In addition, for their default reddening (the histogram
with slanted hatching in Figure~\ref{Fig-PerField}) there would be
many stars even at periods shorter than about 250 days which still
would have $(J-K)_0 > 2.0$ which is not observed in the other
fields. This could off course be real, but it is generally believed
(e.g. Launhardt et al. 2002) that the population of low- and
intermediate mass stars in the Nuclear Bulge (the inner about 30 pc
from the GC) and GB are similar, but that in the former there is an
overabundance of $10^7 - 10^8$ year old stars.  In this picture one
would expect the period distributions to be similar at shorter
periods, essentially independent of latitude. Therefore the period
distribution of stars with $(J-K)_0 < 2.0$ for the increased reddening
is adopted.

The KS test is performed on consecutive fields in latitude for the
distributions based on the stars with $(J-K)_0 < 2.0$.  It is found
that the probability that the distributions are the same is $10^{-10}$ for
the $b = -0.05 /$ $-1.21$ fields, 0.50 for the $b = -1.21 /$ $-1.39$
fields, 0.80 for the $b = -1.39 /$ $-1.81$, and $>0.99$ for the fields
at more negative latitudes. The conclusion is that the period distributions of
the fields at and below $-1.2$ degree are statistically indistinguisable, 
but that the field at $-0.05$ latitude has a significantly different
period distribution (the probability that this distribution is the
same as the distribution of the combined 6 \OG\ fields is $10^{-22}$).
This conclusion is independent of the assumed reddening of the inner
Bulge field, which influences how many stars will have $(J-K)_0 < 2.0$.  
For the default reddening of Schultheis et al., the probability that
the distributions are the same for the $b = -0.05 /$ $-1.21$ fields is
still only 0.0033. The difference in the period distributions is
especially clear at longer periods. Of the 236 stars in the inner
field with $(J-K)_0 < 2.0$ 61 have $P > 500$ days, while in the other
fields this is 3 out of 367.

The difference in period distribution might be due to an under
representation of short period stars in the inner field. However,
Figure~4 in Glass et al. illustrates that the expected $K$-magnitudes
at short periods are not fainter than the completeness limit of their
survey.  In fact, Glass et al. mention that they expect that the
number of short-period Miras ($P < 250$ days) is at least 75\%
complete. As a test, one-third of stars with $P < 250$ days were
randomly duplicated and added to the sample, and the KS test repeated
to find again a large difference between the period distribution of
the field at $-0.05$ degrees and the other fields.

The difference is emphasised in Figure~\ref{Fig-ComparePer} where the
scaled period distribution of stars which have $(J-K)_0 < 2.0$ in the
5 fields between $b = -1.39\degr$ and $b = -5.8\degr$ has been
subtracted from the inner field. The scaling was done in such a way
that at shorter periods the two distributions would cancel at a level
 of 1$\sigma$ (based on Poisson errors). Even if the scaling is done in
a slightly different way the result is always very similar, in the
sense that there is a significant ($>4 \sigma$) overabundance of LPVs
in the inner field bewteen about 350 and 600 days.

The conclusion is that there is a significant population of LPVs with
period $\more 500$ days present in the inner field, which remains
barely present at latitude $-1.2\degr$, and is absent for $b \less -1.4\degr$. 
This was indirectly noted by Glass et al. who noticed that the average
period of the stars in this field at $b = -0.05\degr$ is 427$d$ (and
that of the known OH/IR stars 524$d$), while the average period in the
Sgr {\sc i} window ($b = -2.6\degr$) is 333$d$, with no known OH/IR stars 
(GWCF).

To quantify the nature of the Mira Bulge population, synthetic AGB evolutionary
models have been calculated, which are described in detail in 
Appendix~\ref{AppC}. 

In brief, the synthetic AGB code of Wagenhuber \& Groenewegen (1998)
was finetuned to reproduce the models of Vassiliadis \& Wood (1993)
and then extended to more initial masses and including mass loss on
the RGB. For several initial masses the fundamental mode period
distribution was calculated for stars inside the observed instability
strip and when the mass loss was below a critical value to simulate
the fact that they should be optically visible. Vassiliadis \& Wood
(1993) provide calculations for 4 different metal abundances: $Z =
0.016, 0.008, 0.004$ and 0.001. We used the models for $Z = 0.016$,
representing a solar mix, which are most appropriate for our Bulge
sample (e.g. Rich 1998). We also show that our results are essentially
unchanged if $Z$ = 0.01 or 0.02 are adopted. From the comparison of
the observed period distribution for fields more than $1.2\degr$ away
from the galactic centre with the theoretical ones, we deduce that the
periods can be explained with a population of stars with Main Sequence
masses in the range of 1.5 to 2.0~\msol. A possible extension to
smaller masses is possible, but not necessary to explain the periods
below 200 days.
To explain the excess periods in the range of 350-600 days observed
closer to the centre we need initial masses in the range 2.5 -
3~\msol.  The presence of more massive stars in the inner field at $b$
= $-0.05\degr$ cannot be excluded, as it turns out that for more
massive stars the optically visible Mira phase is essentially
absent. Sevenster (1999) analysed OH/IR stars (which are LPVs with
longer periods and higher mass loss rates than the Miras) in the inner
Galaxy and came to the conclusion that OH/IR stars in the bulge have a
minimum intial mass of about 1.3~\msol, based on an analysis of
infrared colours, compatible with our results. We briefly mention here
the result from Olivier et al. (2001) who studied a sample of LPVs in
the solar neighbourhood with periods in the 300 to 800 days
range. They conclude that majority of these stars had initial masses
in the range of 1 - 2~\msol , with an average value of 1.3~\msol,
lower than what we find for the 300 to 600~days range sample. This
difference may be explained by the fact that our conclusions are only
valid for a sample with no or only low mass loss rates ($\less 5
\times 10^{-6}~\msol/yr$ ), contrary to their sample which was
selected to contain stars with significant mass loss ($\sim
10^{-5}~\msol/yr$).  As can be seen in the Vassiliadis \& Wood (1993)
models, the period increases considerably when the stellar mass is
reduced by the mass loss process.
 
We do not see a variation in the period distributions for the higher
latitude fields (beyond $1\degr$ latitude) and can consider this as a
homogeneous ``bulge'' population, which according to the Vassiliadis \&
Wood (1993) model has ages in the range of 1 to about 3~Gyr.
The excess population closer to the Galactic Centre is younger than 1
Gyr.  According to Launhardt et al. (2002), the Nuclear Bulge
(approximately the central degree) contains besides the bulge population seen
at higher latitudes also an additional population due to recent star
formation closer to the galactic centre. Blommaert et al. (1998) find
that the extrapolation of the number density of bulge OH/IR stars
towards the galactic centre would explain half of the galactic centre
OH/IR population, but that an additional population, intrinsic to the
galactic centre, exists which agrees with what we see in the
distribution here.

The formation history of the Bulge is still a matter of debate. In
several works like in Kuijken \& Rich (2002) and recently in Zoccali
et al. (2004), the bulge is considered to be old ($>10$~Gyr) and
formed on a relatively short timescale ($< 1$~Gyr) (e.g. Ferreras et
al. 2003).  On basis of the modelling of colour-magnitude diagrams,
Zoccali et al. claim that no trace is found of any younger stellar
population than 10~Gyr. The bulge Miras do not fit in this picture as,
according to our analysis, they are considerably younger. The field
studied by Zoccali et al.  is centered at ($l,b$) = (0.277, $-6.167$)
so at a slighter higher latitude than our extreme fields ($b \approx
-5.8\degr$). Although we are limited by the small number of Miras
detected at the highest latitude fields, we do not see a change in
period distribution for those fields.  Zoccali et al. (2004)
acknowledge the presence of Miras, but consider them as part of the
old population.  It is true that Miras are also detected in globular
clusters and thus can be associated with old ages, as is the case for
a 1~\msol\ star in the Vassiliadis \& Wood (1993) model but these
stars produce periods shorter than 200~days (Figure~C.1), insufficient
to explain the period distribution seen in the bulge. The periods
of Miras in Globular Clusters range from 150 to 300 days (Frogel \&
Whitelock 1998) and so only overlap with the shorter periods of the
bulge Miras.

Our results agree more with the analysis of the infrared ISOGAL survey
discussed in van Loon et al. (2003). They conclude that the bulk of
the bulge population is old (more than 7~Gyr) but that a fraction of
the stars is of intermediate age (1 to several Gyr).  The Miras in our
study can thus be considered as the intermediate age population seen
in their analysis. van Loon et al.  also see evidence for an even
younger population ($< 200$~Myr), but according to our findings, this
would be restricted to the area close to the Galactic Centre.

Our discussion on the ages of the Mira stars is based on the
assumption that they have evolved from single stars. An alternative
scenario as suggested by Renzini \& Greggio (1990), would be that the
brighter (longer period) Miras could evolve from close binaries where
the components coalesced to form one single star. This could lead to
an underestimation of the age as the Mira essentially is the product
of lower mass and thus older stars. This scenario may seem in better
agreement with the idea that the bulge consists of an old stellar
population. It however suffers from the same problem as the
intermediate age population, in the sense that also no clear evidence
for Blue Stragglers (which would be the Main Sequence counterpart of
the Miras) is found (Kuijken \& Rich, 2002).

If indeed the bulk of the bulge population is old and formed quickly
and if the Miras are of intermediate age, then our Miras must be
representatives of a population which was added at a later stage and
it is unclear how it relates to the overall bulge.
An interesting scenario suggested in Kormendy \& Kennicutt (2004) is
the one in which a secondary bulge or also called pseudo-bulge forms
within an old bulge. Such a process would be connected to the presence
of a ``bar'' which would add ``disky'' material into the old classical
bulge.  The Miras are indeed situated in a bar-structure as is
discussed in the following section.

\begin{figure*}

\begin{minipage}{0.48\textwidth}
\resizebox{\hsize}{!}{\includegraphics{PerDistGLASS.ps}}
\end{minipage}

\begin{minipage}{0.48\textwidth}
\resizebox{\hsize}{!}{\includegraphics{PerDistbul44_pap.ps}}
\end{minipage}
\hfill
\begin{minipage}{0.48\textwidth}
\resizebox{\hsize}{!}{\includegraphics{PerDistbul03_pap.ps}}
\end{minipage}

\begin{minipage}{0.48\textwidth}
\resizebox{\hsize}{!}{\includegraphics{PerDistbul05_pap.ps}}
\end{minipage}
\hfill
\begin{minipage}{0.48\textwidth}
\resizebox{\hsize}{!}{\includegraphics{PerDistbul22_pap.ps}}
\end{minipage}

\begin{minipage}{0.48\textwidth}
\resizebox{\hsize}{!}{\includegraphics{PerDistbul37_pap.ps}}
\end{minipage}
\hfill
\begin{minipage}{0.48\textwidth}
\resizebox{\hsize}{!}{\includegraphics{PerDistbul06_pap.ps}}
\end{minipage}

\caption[]{
Mira period distribution for 7 fields with similar longitudes but a
range in latitudes (as indicated in the top right corner). For the
field at $b \sim -5.8\degr$, \OG\ fields 6 and 7 have been combined.
For the shaded histograms only stars with $(J-K)_0 < 2.0$ have been
included.  The field at $(-0.05, -0.05)$ is based on Glass et
al. (2001), see main text for details. The histogram with slanted
hatching is for the reddening by Schultheis et al. (1999b) for stars
in this field, the shaded histogram for the adopted reddening which is
1.35 times larger.
}
\label{Fig-PerField}
\end{figure*}

\begin{figure}
\centerline{\psfig{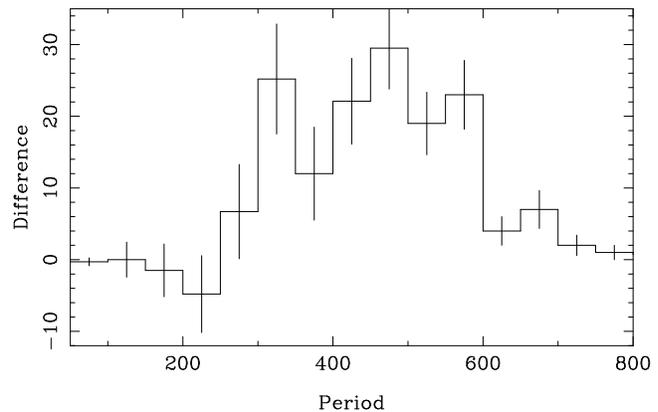}}
\caption[]{
Period distribution of the stars in the $b = -0.05\degr$ field with
$(J-K)_0 < 2.0$ minus the suitably scaled period distribution of the
stars in the $b = -1.39$ to $-5.8\degr$ fields with $(J-K)_0 < 2.0$. The
scaling is such that below \less 300 days the two distributions 
cancel at the 1$\sigma$ level.
}
\label{Fig-ComparePer}
\end{figure}

\section{The orientation of the Bar}
\label{sect-orien}

Table~\ref{tab-ZP} lists the zero points (ZPs) of the the $K$-band
$PL$-relation (for a fixed slope of $-3.37$) for the \OG\ fields
individually. To increase the statistics, some neighbouring fields
have been added together, as indicated in the first column of the table. The
galactic coordinates listed are the mean values of all individual
objects, rather than the mean of the field centres.
Figure~\ref{Fig-ZP} plots these ZPs (with error bars) as a
function of Galactic longitude. There is a clear correlation; the
formal weighted fit has a slope of $-0.023 \pm 0.005$
(magnitude/degree).  Restricting the fields to those with longitudes
$-5 < l < +5$ (reducing the contamination by disk stars, see 
Appendix~\ref{AppB}) the fit becomes:
\begin{equation}
m_{\rm K}= (-0.0192 \pm 0.0087)\;\; l \; + \; (15.484 \pm 0.019)
\end{equation}
with an rms of 0.10 and based on 32 fields.

The interpretation of this correlation is that the Bulge Miras are
located in the Galactic Bar that has a certain orientation towards the
observer. A similar correlation was found by Wray (2004) who concluded
that an appropriately chosen ZP in $I$ for the small amplitude
\OG\ variables in their sample (which they identify as to correspond
to in Box ``A-'') correlated with Galactic longitude. No estimate for
the orientation of the Bar was given however.

\begin{figure}
\centerline{\psfig{figure=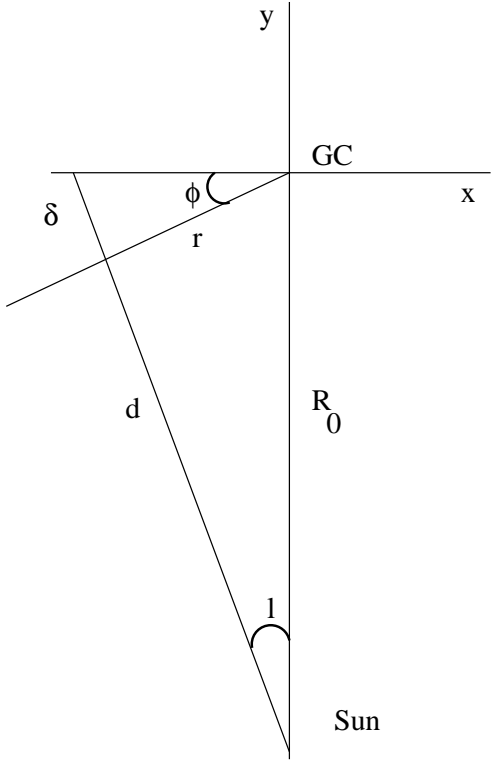,width=6.5cm}}
\caption[]{
Schematic drawing of the orientation of the major-axis of the Galactic
Bar w.r.t. the Sun and the Galactic Centre. The $z$-axis is directed
towards the reader. See Appendix~\ref{AppB}. }
\label{Fig-ORI}
\end{figure}

\begin{figure}
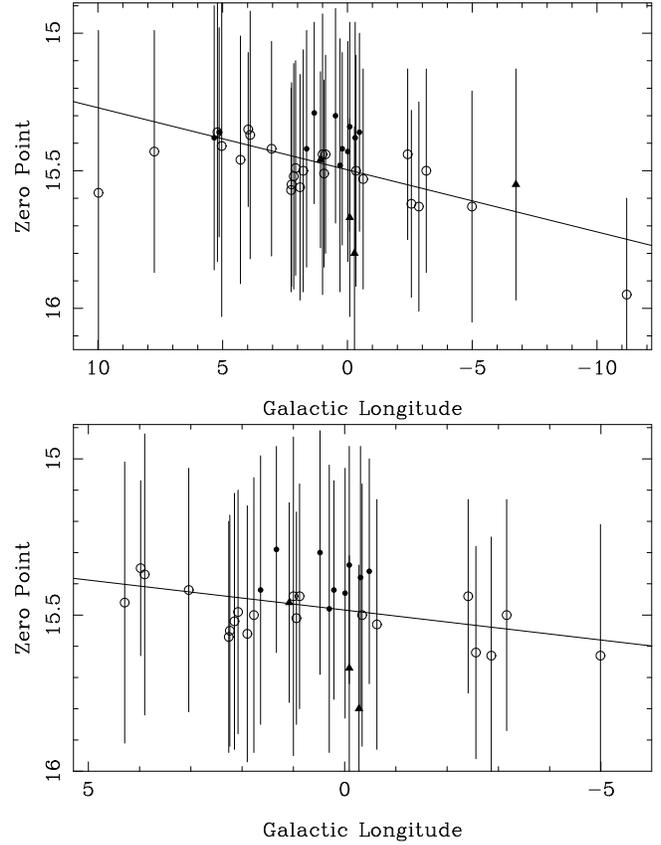

\includegraphics[width=85mm]{ZP_0.ps}

\includegraphics[width=85mm]{ZP_0_-5to+5.ps}
\caption[]{
Zero point of the $K$-band $PL$-relation as a function of Galactic
longitude. Galactic latitudes below $-4.0$ are indicated by filled
triangles, those larger than $-2.6$ by filled circles, and the
remaining by open circles. Error bars are also plotted. The lines
represent linear least-squares fits to all fields (top panel), and
those with $\mid l \mid <5\degr$ (bottom panel). }
\label{Fig-ZP}
\end{figure}

In Appendix~\ref{AppB} Monte Carlo simulations are carried out in order
to quantify two issues: can these observations be used to constrain
the orientation of the Galactic Bar, and, second, given the specific
location of the \OG\ fields, if there is any bias in the derived zero
point compared to a fiducial ZP when all Miras would be located
exactly in the Galactic Centre (GC). As described in Appendix~\ref{AppB}, 
for a spatial distribution of Bulge and Disk stars following Binney et
al. (1997), viewing angles $\phi$ of 43 and 79 degrees (see the 
orientation in Figure~\ref{Fig-ORI}) result in slopes (magnitude
versus $l$, Eq.~2) in agreement with observations. However, the model with
$\phi = 43\degr$ gives a much better fit to the number of stars per
field. The bias in the ZPs is essentially independent of viewing
angles, and for the best fitting model the observed ZP derived for all
stars (Eq.~1) is too bright by 0.018 mag ($\pm$ 0.013), while the ZP in
Eq.~2 is too bright by 0.002 ($\pm$ 0.021) mag.

The preferred value of $\phi = 43\degr$  is in agreement with the
values of about $45\degr$ by Whitelock (1992), based on 104 IRAS detected
Mira variables, and the preferred value of $46\degr$ by Sevenster et al. (1999),
based on an analysis of OH/IR stars in the inner Galaxy. 

Other values in the literature are usually much larger, between 60 and
80 degrees: Dwek et al. (1995) and Binney et al. (1997), based on COBE-DIRBE data, 
Stanek et al. (1997), based on bulge red clump stars, Robin et al. (2003) and 
Picaud \& Robin, based on colour-magnitude fitting. 
Sevenster et al. (1999), however,  argues that these values 
are commonly found when no velocity data is available, the longitude
range is too narrow or when low latitudes are excluded.
It is also possible that these studies are tracing other populations, which
may be differently distributed than the Miras. Whitelock et al.
and Sevenster et al. do use populations closely related to the Mira stars and find 
an angle of the bar close to the one we derive.

\begin{table*}
\caption{Zero point of the $K$-band $PL$-relation. }
\begin{tabular}{rrrrrr} \hline
Fields    & $l^{(a)}$ & $b^{(a)}$  & ZP$^{(b)}$  & N$^{(c)}$    &  \\ 
          & (deg) & (deg)   & (mag)              &      & \\ \hline
8,9,10,11 & 9.98  & $-3.75$ & 15.58 $\pm$   0.59 &   23 & \\
12,13     & 7.74  & $-3.50$ & 15.43 $\pm$   0.44 &   27 & \\
15        & 5.34  & $+2.59$ & 15.38 $\pm$   0.48 &   22 & \\
17        & 5.21  & $-3.50$ & 15.36 $\pm$   0.47 &   24 & \\
14        & 5.14  & $+2.73$ & 15.36 $\pm$   0.38 &   29 & \\
16        & 5.04  & $-3.37$ & 15.41 $\pm$   0.62 &   17 & \\
42        & 4.29  & $-3.52$ & 15.46 $\pm$   0.45 &   13 & \\
19        & 3.98  & $-3.41$ & 15.35 $\pm$   0.28 &   20 & \\
18        & 3.90  & $-3.19$ & 15.37 $\pm$   0.45 &   20 & \\
35,36     & 3.04  & $-3.18$ & 15.42 $\pm$   0.39 &   29 & \\
33        & 2.24  & $-3.72$ & 15.55 $\pm$   0.37 &   22 & \\
32        & 2.26  & $-3.22$ & 15.57 $\pm$   0.37 &   24 & \\
31        & 2.15  & $-3.01$ & 15.52 $\pm$   0.41 &   32 & \\
 2        & 2.08  & $-3.59$ & 15.49 $\pm$   0.39 &   19 & \\
30        & 1.90  & $-2.90$ & 15.56 $\pm$   0.41 &   25 & \\
21        & 1.77  & $-2.70$ & 15.50 $\pm$   0.44 &   33 & \\
20        & 1.64  & $-2.52$ & 15.42 $\pm$   0.43 &   41 & \\
34        & 1.33  & $-2.43$ & 15.29 $\pm$   0.33 &   35 & \\
46        & 1.08  & $-4.14$ & 15.46 $\pm$   0.32 &   18 & \\
 1        & 0.94  & $-3.69$ & 15.51 $\pm$   0.34 &   22 & \\
45        & 1.00  & $-3.96$ & 15.44 $\pm$   0.51 &   15 & \\
38        & 0.88  & $-3.49$ & 15.44 $\pm$   0.36 &   25 & \\
39        & 0.48  & $-2.26$ & 15.30 $\pm$   0.39 &   69 & \\
 4        & 0.30  & $-2.10$ & 15.48 $\pm$   0.46 &   54 & \\
43        & 0.21  & $+2.84$ & 15.42 $\pm$   0.35 &   47 & \\
 3        & -0.01 & $-2.02$ & 15.43 $\pm$   0.40 &   70 & \\
37        & -0.09 & $-1.81$ & 15.34 $\pm$   0.38 &   68 & \\
 7        & -0.09 & $-5.86$ & 15.67 $\pm$   0.36 &    6 & \\
 5        & -0.31 & $-1.39$ & 15.38 $\pm$   0.42 &   90 & \\
 6        & -0.28 & $-5.70$ & 15.80 $\pm$   0.46 &    8 & \\
22        & -0.34 & $-3.01$ & 15.50 $\pm$   0.42 &   46 & \\
44        & -0.48 & $-1.26$ & 15.36 $\pm$   0.36 &   79 & \\
23        & -0.63 & $-3.44$ & 15.53 $\pm$   0.40 &   30 & \\
25        & -2.41 & $-3.64$ & 15.44 $\pm$   0.31 &   33 & \\
24        & -2.56 & $-3.46$ & 15.62 $\pm$   0.34 &   29 & \\
41        & -2.86 & $-3.35$ & 15.63 $\pm$   0.38 &   24 & \\
40        & -3.16 & $-3.27$ & 15.50 $\pm$   0.37 &   34 & \\
26,27     & -4.99 & $-3.58$ & 15.63 $\pm$   0.42 &   39 & \\
28,29     & -6.75 & $-4.55$ & 15.55 $\pm$   0.42 &   13 & \\
47,48,49  & -11.19 & $-2.78$ & 15.95 $\pm$  0.35 &   18 & \\
\hline
\end{tabular}
\label{tab-ZP}

(a) Mean values of individual objects

(b) Zero point of the $K$-band $PL$-relation for a slope of $-3.37$.

(c) Number of objects in Box ``C'' used to calculate the $PL$-relation.

\end{table*}

\section{The distance to the Galactic Centre}

A slope of the $K$-band $PL$-relation of $-3.37 \pm 0.09$ is derived. 
There appear not to exist many previous determination of this quantity. 
Recently, GS03 derived a $PL$-relation in NGC 6522 based on 34 MACHO
variables with $r$-amplitude $>$1.5 and \DE\ $K$ photometry: 
$m_{\rm K} = -4.6 \; \log P + 18.1$. No errors or rms were given,
as--by their own account--this fit was made by eye.  Much better
agreement is found with GWCF. Based on multi-epoch data of 55 stars
they found $m_{\rm K} = (-3.47 \pm 0.35)\; \log P + (15.64 \pm 0.86)$
(rms=0.35) in the Sgr {\sc i} field.

Zero points for the $K$-band $PL$-relation have been derived in two ways. 
First, a direct fit to all stars resulting in (15.44 $\pm$ 0.21),
and secondly, determining ZPs per (sub)-field, and fitting this as a
function of $l$, resulting in (15.484 $\pm$ 0.019).

Applying the small bias corrections discussed at the end of
Sect.~\ref{sect-orien} and averaging over the two estimates, the
adopted $K$-band $PL$-relation for Miras at the GC is:
\begin{equation}
m_{\rm K}= -3.37  \log P + (15.47 \pm 0.03)
\end{equation}

The derived $PL$-relation can be compared to the one derived for 83
O-rich LPVs in the LMC derived in G04: $m_{\rm K}= (-3.52 \pm 0.16)\;
\log P + (19.56 \pm 0.38)$, with an rms of 0.26. Since the slopes are
not exactly the same, the magnitudes are compared at the approximate
mean period of $\log P = 2.45$. The difference in magnitude is 3.72.
Adopting the LMC based slope of $-3.52$ for the GB Miras, and re-fitting the ZP,
the bias corrected ZP would become 15.85, resulting in a GB-LMC
DM difference of 3.71, essentially the same value.
If the distance to the GC is assumed to be 7.94 kpc (Eisenhauer et
al. 2003; in a recent preprint this was even lowered to 7.62
$\pm$ 0.32 kpc, Eisenhauer et al. (2005)), then the LMC would be at a
DM = 18.21, or if the DM to the LMC is assumed to be 18.50 (Walker
2003, Feast 2004a), then the GC would be at 9.0 kpc. A similar result
was found by GWCF who derived a distance to the GC of 8.9 $\pm$ 0.7
kpc, assuming 18.55 for the LMC DM and $\phi = 45\degr$.
The analysis so far has assumed no metallicity dependence of the
Mira $PL$-relation. Wood (1990) present linear non-adiabatic pulsation
calculations that suggest a dependence of the form $\log P \sim 0.46
\log Z +1.59 \log L$, but he notes that in the $K$-band the dependence
is expected to be weaker and following the example he presents one
infers a dependence of $0.25 \log Z$ in the $K$-band. 
In G04 $K$-band $PL$-relations were derived for carbon-miras in the
SMC and LMC. At a characteristic period of $\log P = 2.45$ one infers
a relative difference in DM of 0.38, which is smaller than the
commonly quoted value of near 0.50 (0.48-0.53 $\pm$ 0.11, FO cepheids
[Bono et al. 2002], 0.46-0.51 $\pm$ 0.15, FU cepheids [Groenewegen 2000], 
0.44 $\pm$ 0.05, TRGB [Cioni et al. 2000]). This may hint at a
metallicity dependence of the Mira $K$-band $PL$-relation.
To test this hypothesis, a correction to the $K$-magnitude of $+\beta
\log Z$ will be assumed\footnote{that is, $M_{\rm K} = M_{\rm K}({\rm
ref}) + \beta \log (Z/Z_{\rm ref})$, where $M_{\rm K}({\rm ref})$ is
the known magnitude is a galaxy with metallicity $Z_{\rm ref}$, and
$M_{\rm K}$ the magnitude it would have in a galaxy with metallicity
$Z$} (for both O- and C-rich LPVs), and the Bulge, LMC, and SMC will
be assumed to have solar, solar/2 and solar/4 metallicity,
respectively.  For a value $\beta = 0.25$ the relative SMC-LMC DM
based on the C-Miras is increased from 0.38 to 0.46, while the
relative DM LMC-GC is increased from 3.72 to 3.80. If the relative
SMC-LMC DM is fixed at 0.50, then $\beta = 0.40$ is required, and the
relative DM LMC-GC becomes 3.84 for that value. For a LMC DM of 18.50,
the distance to the GC then becomes 8.6 kpc. The error in this value
is somewhat difficult to estimate as the $PL$-relations derived in G04
and here are from--at best--the average of two $K$ values. Work by
Feast et al. (1989) indicates that in the case of multi-epoch data
(and for the small depth effect in the LMC) the intrinsic dispersion in
the $PL$-relation is about 0.13 mag. Therefore we assign an error of
0.18 to the difference in DM. This implies an error of 0.7 kpc.
Based on this large sample of Mira variables in the direction of the
GB the conclusion is that the distance to the GC is between 8.6 and
9.0 ($\pm 0.7$) kpc, depending on the metallicity dependence of the
$K$-band $PL$-relation.
Feast (2004b) discusses the zeropoint of the Mira $K$-band
$PL$-relation, and adopting the slope observed in the LMC ($-3.47$)
derives a zeropoint of 1.00 $\pm$ 0.08, averaging over independently
derived ZPs from trigonometric parallaxes, OH VLBI expansion
parallaxes and Galactic Globular Clusters.
Adopting a slope of $-3.47$ and refitting the ZP of the Bulge sample,
the bias corrected value becomes 15.73 $\pm$ 0.03, and without
metallicity correction (consistent with the assumption above about the
metallicities in Bulge, LMC, SMC) leads to a distance to the GC of 
8.8 $\pm$ 0.4 kpc.  This independent distance estimate is in between
the values derived using no or a strong metallicity dependent zero point.

\acknowledgements{

This research has made use of the SIMBAD database, operated at CDS,
Strasbourg, France.

This publication makes use of data products from the Two Micron All
Sky Survey, which is a joint project of the University of
Massachusetts and the Infrared Processing and Analysis
Center/California Institute of Technology, funded by the National
Aeronautics and Space Administration and the National Science Foundation.

}

{}

\appendix

\section{The light curve analysis model}
\label{AppA}

Some small changes w.r.t. the implementation of the lightcurve
analysise model in G04 are described.

The only change in the first part of the code is the level at which a
period is accepted as significant. This level was set at {\it
significance = $ 1.0 \times 10^{-16}$}, compared to $5.5 \times
10^{-11}$ in G04. This was possible as--contrary to G04--only objects
with large amplitudes were searched for. The resulting increase in the
number of spurious periods was then caught in the process of visual
inspection.

As in G04 a list of known objects (both known non-LPVs, and known AGB
giants and LPVs) was compiled to ease automatic association. The list
comprises:

(1) 14833 IRAS sources within 10 deg. radius of the centre of the 49
\OG\ fields at RA = 268.87, Dec = $-31.03$\footnote{Retrieved from the
  infrared science archive at IPAC, http://irsa.ipac.caltech.edu/},

(2) 51141 ISOGAL sources from Omont et al. (2003, hereafter OGA03;
those within the extreme values of the \OG\ field boundaries 261.612
$<=$ RA $<=$ 276.133, $-40.726 <=$ Dec $<= -21.328$),

(3) 268 pulsating variable stars, 1650 Eclipsing Binaries, and 943
Miscellaneous variable stars from {\sc ogle-i} (Udalski et al. 1994,
1995a, 1995b, 1996, 1997),

(4) 332 objects from Alard et al. (2001, hereafter ABC01) who
correlated ISOGAL sources within the NGC 6522 and Sgr {\sc I} Baade
windows with the MACHO database,

(5) 2353 objects from Ojha et al. (2003, hereafter OOS03) who studied
sources in 9 ISOGAL fields,

(6) 174 M-giants later than spectral type M0 in the NGC 6522 Baade's
window from Glass \& Schultheis (2002, hereafter GS02),

(7) 421 objects from Glass et al. (2001, and erratum), who monitored
in $K$ over four years an 24x24 arcmin area near the Galactic Centre,

(8) 122 objects from Alard et al. (1996, identified as Ter-[number])
who identified LPV using red photographic plates,

(9) 494 objects from Lloyd-Evans (1976, hereafter  Lloyd-Evans (1976) who identified
Mira variables in the three Baade windows, identified as
TLE-[field]-[number] in Table~\ref{TAB-B}), Blanco et al. (1984;
herafter BMB, who identified M giants in Baade's window), and Blanco
(1984; herafter B84, who identified RR Lyrae variables) with
coordinates listed in the {\sc simbad} database to 1\arcsec\ or
better,

(10) 33 Nova related objects from Cieslinski (2003).  The total number
of sources used in the automatic correlation is 72764.

\section{Simulations of the Galactic Bulge and foreground disk stars}
\label{AppB}

\begin{figure}
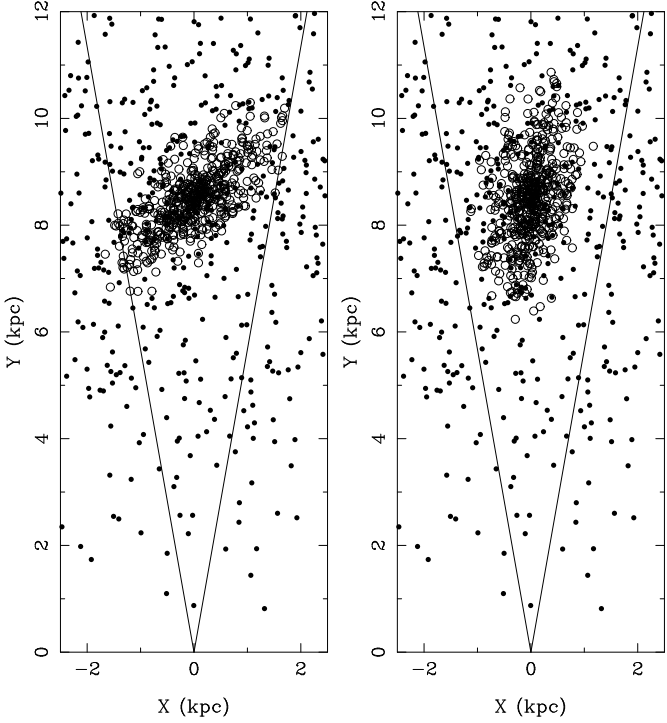


\begin{minipage}{0.24\textwidth}
\resizebox{\hsize}{!}{\includegraphics{barsim_43.ps}}
\end{minipage}
\hfill
\begin{minipage}{0.24\textwidth}
\resizebox{\hsize}{!}{\includegraphics{barsim_79.ps}}
\end{minipage}
\caption[]{
For angles of $\phi$ of 43 and 79 degrees (left and right panel), the
projection (drawn to scale) on the Galactic Plane of 1300 randomly
drawn stars. Disk stars are represented by filled circles, Bulge stars
by open circles.  The Sun is at X= 0, Y= 0. The Galactic Centre is at  X= 0, Y= 8.5. 
The lines illustrate Galactic longitudes of $\pm$10 degrees.
}
\label{Fig-Proj}
\end{figure}

In this Appendix the calculations are described to model a population
in the direction of the Galactic Bulge.

The basic model is essentially the one proposed by Binney et al. (1997) 
to model the dust-corrected near-infrared COBE/DIRBE surface
brightness map of the inner galaxy. The number density of bulge stars is assumed to be:

\begin{equation}
f_{\rm b} = f_0 \exp(-a^2/a_{\rm m}^2) \; / \; (1 + a / a_0)^\beta
\end{equation}
with $f_0 = 624$, $a_{\rm m} = 1.9$ kpc, $a_0 = 0.10$ kpc, $\beta = 1.8$. 
Binney et al. assumed a tri-axial bulge with axial ratios 1 : 0.6 : 0.4. 
For numerical convenience a prolate ellipsoid is assumed here: 
$a = \sqrt{ x^2 + (y/ \eta)^2 + (z/ \eta)^2}$ with the value of $\eta = 0.5$ taken from Binney et al.

The number of Bulge objects up to a radius $r$ from the centre, that
defines the probability density function in the simulation, is approximated as:
\begin{equation}
N_{\rm b}(r) =  \int_{0}^{r} 4 \pi a^2 f_{\rm b}(a) \; da
\end{equation}
up to a maximum radius that is taken to be the co-rotation radius,
with a default value of $R_{\rm cr} = 2.4$ kpc, following Dwek et al. (1995).

The number density of disk stars is assumed to be:

\begin{displaymath}
f_{\rm d} = \left(\exp(-\mid z\mid/ z_0) + \alpha \exp(-\mid z\mid/ z_1)\right) \; \times
\end{displaymath}
\begin{equation}
\hspace{20mm}   R_{\rm d} \;  (\exp(-r/R_{\rm d}) - f_{\rm h} \; \exp(-r/R_{\rm h})) 
\end{equation}
with $z_0 = 0.210$ kpc, $z_1 = 0.042$ kpc, $\alpha = 0.27$, $R_{\rm d} = 2.5$ kpc (Binney et al.) 
and $R_{\rm h} = 1.3$ kpc (Picaud \& Robin 2004). This functional form
follows Binney et al., but also allows for a ``hole'' in inner disk
(the scaling parameter is 0 $\leq f_{\rm h} \leq 1$, and identical to
zero in Binney et al.). The total number of disk stars, and the
probability density functions, are defined as:

\begin{displaymath}
N_{\rm d} \equiv  N_{\rm d,z}(z) \times N_{\rm d,r}(r)
\end{displaymath}
given by,

\begin{displaymath}
N_{\rm d} =  \left[ 2 \int_{0}^{z} \exp(-z / z_0) + \alpha \; \exp(-z / z_1) \; dz\right] \; \times
\end{displaymath}
\begin{equation}
       \hspace{10mm}  \left[ \int_{0}^{r} 2 \pi r R_{\rm d}  (\exp(-r/R_{\rm d})- f_{\rm h} \; \exp(-r/R_{\rm h}))  \; dr \right]
\end{equation}
up to maximum values $z_{\rm max} = (R_{\rm cr} \; \eta)$, 
and $R_{\rm max} = 8.0$ kpc, respectively.

A disk or bulge star is generated according to the ratio $N_{\rm b} /
(N_{\rm b} + N_{\rm d,z} \times N_{\rm d,r})$. In case of a disk
star, its height above the plane, $z$, distance to the GC, $r$, and
a random angle between 0 and $2 \pi$ in the Galactic plane, are drawn
according to the probability functions $N_{\rm d,z}(z)$ and $N_{\rm d,r}(r)$. 
Its coordinates $x,y,z$ are then known.

In case of a bulge star, the distance, $a$, to the GC is drawn
according to the probability function $N_{\rm b}$, and then a star is
randomly placed on the surface of the appropriate ellipsoid, to find
$x,y,z$. These values are then rotated by an angle $\phi$ in the
Galactic plane (see Fig.~\ref{Fig-ORI}). 

The Galactic coordinates are then derived assuming a distance from the
Sun to the GC of $R_0 = 8.5$ kpc, and height above the plane of 
$z_0 = +24$ pc (Ma\'iz-Apell\'aniz 2001)

In a second step, for every star, the known distance to the Sun is
used to calculate its appararent magnitude, assuming an arbitrary
$M_{\rm K}$ of $-7.5$ mag with a Gaussian dispersion of ${\sigma}_{\rm K} = 0.15$ mag. 
This is about the dispersion observed in the $PL$-relation in LMC
Miras when multi-epoch photometry is available to accurately determine
mean-light magnitudes (Feast et al. 1989).

In a third step, for every simulated star it is verified if it is
located within one of the 40 lines-of-sight considered, listed in
Table~\ref{tab-ZP}.  The field sizes of 14.2\arcmin$\times$57\arcmin\
are approximated by a circle of radius 0.27 degrees. If so, it is
assumed the star would have been ``detected'' (Given the relative
brightness of the LPVs it is assumed that completeness is not an issue).

The number of stars drawn is such that a total of about 1200 objects
is ``detected'', similar to the actual number. At the end of the
simulation, the average magnitude and dispersion per line-of-sight is
determined, and a weighted least-square fit is made of the mean
magnitude versus longitude, for all fields, and for those with $\mid l
\mid <5\degr$, as for the observations. In addition, the mean
magnitude and dispersion for all ``detected'' stars is determined.

Such a simulation is repeated 1000 times. Then, distribution functions
and from that median and 1-sigma values of the following parameters
are determined: the number of stars (total, disk, bulge), mean magnitude
and sigma (for every line-of-sight), mean magnitude and sigma for all
stars, and slope and error in the slope both when fitted over all
longitudes, and for those fields with $\mid l \mid <5\degr$.

For the standard model of Binney et al. described above (i.e. $f_{\rm h} = 0$) 
it turns out that for two values of $\phi$ a slope (because of the
contamination by disk stars in the outer fields, the slope fitted over
$\mid l \mid <5\degr$ is considered for now on) in agreement with
observations is found: $\phi = 43$ and $79$ degrees (with values
between 25 and 85 degrees resulting in predicted slopes within
1-$\sigma$ of the observed one.)

\begin{figure*}
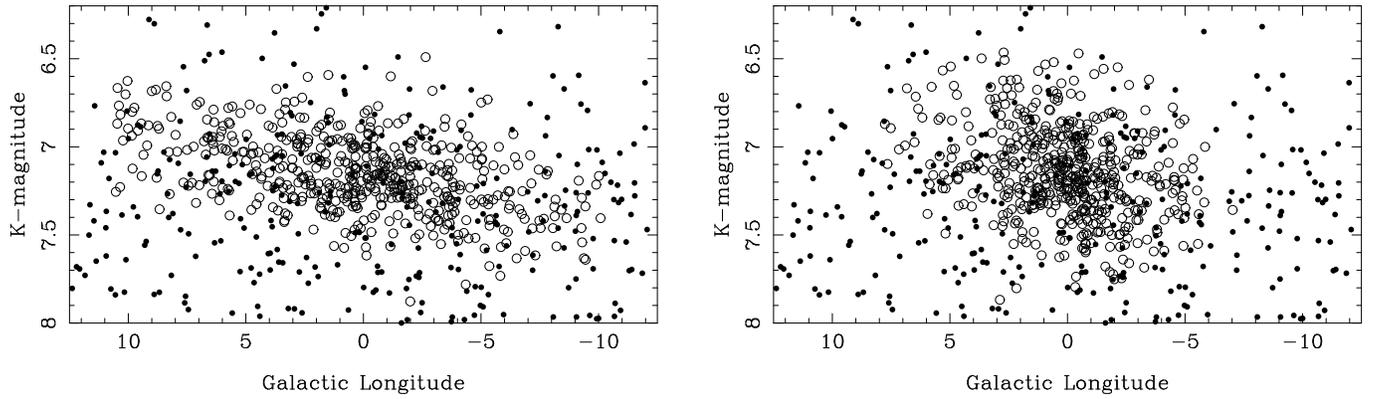


\begin{minipage}{0.48\textwidth}
\resizebox{\hsize}{!}{\includegraphics{barsim1_43.ps}}
\end{minipage}
\hfill
\begin{minipage}{0.48\textwidth}
\resizebox{\hsize}{!}{\includegraphics{barsim1_79.ps}}
\end{minipage}
\caption[]{
For angles of $\phi$ of 43 and 79 degrees (left and right panel), the
distribution of $K$-magnitudes as a function of Galactic longitude for
the stars shown in Fig.~\ref{Fig-Proj}. Disk stars are represented by
filled circles, Bulge stars by open circles.
}
\label{Fig-KMAG}
\end{figure*}

\begin{figure*}
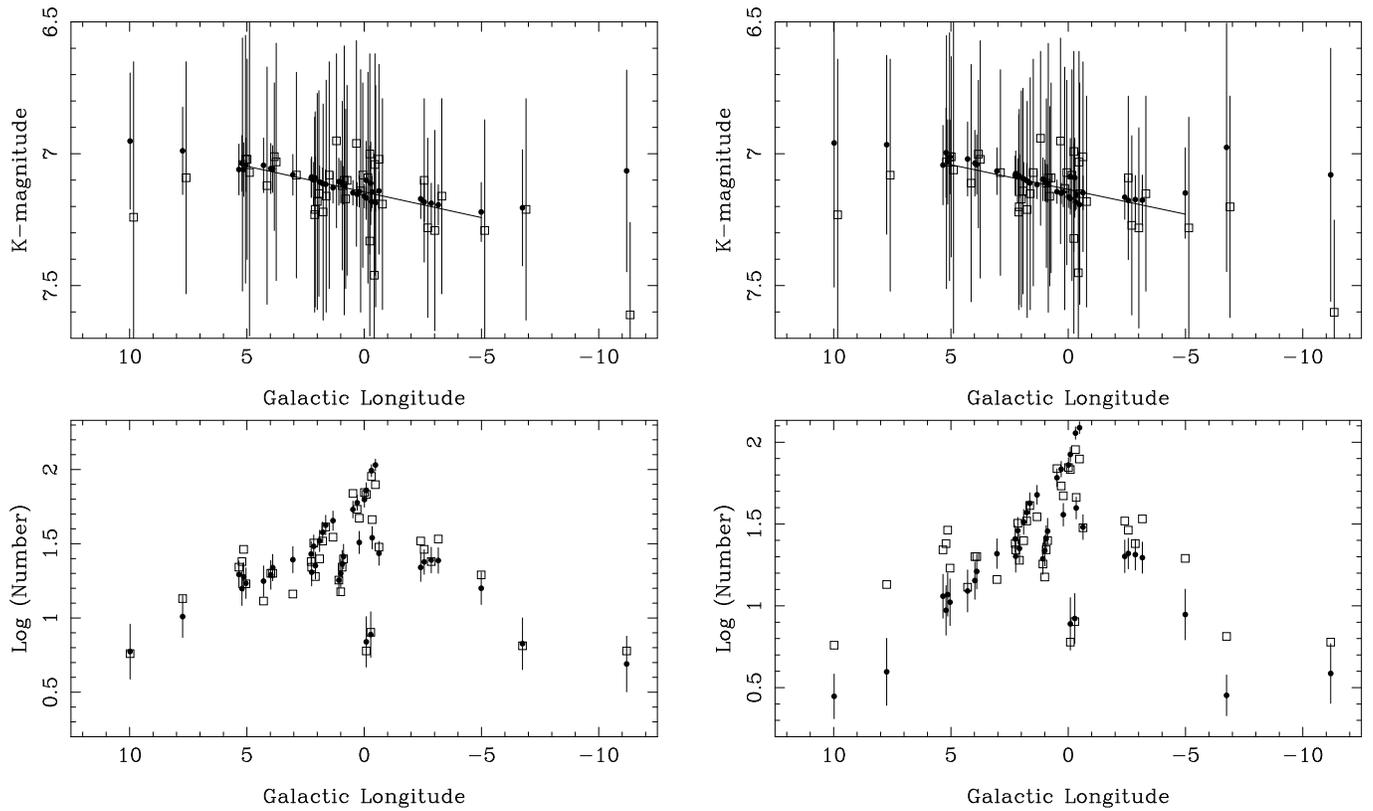


\begin{minipage}{0.48\textwidth}
\resizebox{\hsize}{!}{\includegraphics{anabarsim_43.ps}}
\end{minipage}
\hfill
\begin{minipage}{0.48\textwidth}
\resizebox{\hsize}{!}{\includegraphics{anabarsim_79.ps}}
\end{minipage}

\caption[]{
For angles of $\phi$ of 43 and 79 degrees (left and right panel), the
top panel compares the observed (open squares) and modelled (filled
circles) mean magnitudes and errors for the 40 fields listed
Table~\ref{tab-ZP}, with the line indicating the best fit over the
range $\mid l \mid <5\degr$. For clarity the observed and modelled points
are slightly off-setted in longitude.
The bottom panel compares the observed (open squares) and modelled
(filled circles) number of stars per field.
For both angles the observed slope over the range $\mid l \mid <5\degr$ is
reproduced, but the distribution of stars is much better fitted for $\phi = 43\degr$.
}
\label{Fig-SLOPES}
\end{figure*}

Figures~\ref{Fig-Proj}, \ref{Fig-KMAG} and \ref{Fig-SLOPES} show the
results for these two cases. Figure~\ref{Fig-Proj} shows the
distribution on the Galactic Plane for a random sub-sample of all
stars simulated, and illustrates a fundamental difference between the
two cases. For large viewing angles the outer fields $\mid l \mid
\more 10\degr$ will be dominated by disk stars.
Figure~\ref{Fig-KMAG} shows for the same random sub-sample the observed magnitude as a function of $l$.
In Figure~\ref{Fig-SLOPES} the simulated mean magnitude and error for
each field are compared to the observations in the top panel, while in
the bottom panel the observed and predicted number of objects are compared.
It is from this plot that one may conclude that the model with $\phi = 43\degr$ 
is to be preferred over the one with $\phi = 79\degr$ as the latter model
predicts too few stars, especially in the outer fields.
Comparing only the observed and predicted number of stars (in a
$\chi^2$ sense) a best fit is found for $\phi = 35\degr$ (with values
between 0 and 60 degrees resulting in a reduced $\chi^2$ within 1 unit
of the minimum $\chi^2$). Combining the constraints from the slope and
the number of stars a viewing angle of $\phi = 43 \pm 17$ degrees is
the preferred value.

One may consider the ratio of Bulge to Disk stars as uncertain, and
therefore, a model was considered with $\phi = 79\degr$ and $f_0 = 350$.
The latter value was set so that the model predicted the observed
number of stars in the $l = -11\degr$ field. Such a model would still
underestimate the number of stars in the outer fields at positive $l$,
and would give a slope no longer in agreement with observations.

Finally, a model including a hole in the inner disk is considered
(i.e. $f_{\rm h} = 1$). To have the same ratio of bulge to disk stars,
$f_0$ was set to 425. The results are very similar and the best
fitting angle is now 40 degrees.

For reference, the predicted number of disk and bulge stars for the
viewing angles of 43 and 79 degrees, and for the model with $\phi = 40\degr$ 
and the central hole in the disk, are listed in Tab.~\ref{tab-MODEL}.
As they are quite different, these predictions may be usefull when
additional data (proper motion\footnote{Sumi et al. (2004) present
proper motions for 5 million objects in the \OG\ fields, centered on
the expected $I_0$ magnitude and $(V-I)_0$ colour of red clump giants
but also including some red giants. Of the 2691 LPVs in the present
study, 1612 are listed in Sumi et al.}, radial velocities) become
available to constrain the ratio of disk to bulge objects as a
function of galactic coordinates.

\begin{table*}
\caption{Observed total, and predicted number of disk and bulge stars per field.}
\begin{tabular}{rrrrrrrrr} \hline
     &         &               & \multicolumn{2}{c}{$\phi = 43\degr$}  & \multicolumn{2}{c}{$\phi = 79\degr$} 
                                                                           & \multicolumn{2}{c}{$\phi = 40\degr, f_{\rm h} = 1$}  \\
 $l$ &  $b$    & $N_{\rm obs}$ & $N_{\rm disk}$ & $N_{\rm bulge}$ &  $N_{\rm disk}$ & $N_{\rm bulge}$ &  $N_{\rm disk}$ & $N_{\rm bulge}$ \\ 
\hline
9.98 & $-3.75$ & 5.75  & 2.1 $\pm$ 1.5  &  3.1 $\pm$  2.0  &  2.8 $\pm$ 0.9  &  0.0 $\pm$  0.0     & 2.0 $\pm$ 1.5  & 3.9 $\pm$ 1.9 \\
7.74 & $-3.50$ & 13.5  & 3.1 $\pm$ 1.6  &  6.8 $\pm$  2.6  &  2.8 $\pm$ 1.6  &  1.8 $\pm$  1.2     & 3.0 $\pm$ 1.7  & 7.5 $\pm$ 2.8 \\
5.34 & $+2.59$ & 22  & 5.6 $\pm$ 2.4  &  14.0 $\pm$  3.5  &  4.0 $\pm$ 1.9  &  7.4 $\pm$  2.6     & 4.2 $\pm$ 2.2  & 15.4 $\pm$ 3.9 \\
5.21 & $-3.50$ & 24  & 3.8 $\pm$ 1.7  &  12.0 $\pm$  3.6  &  2.9 $\pm$ 1.7  &  6.6 $\pm$  2.5     & 2.9 $\pm$ 1.7  & 12.9 $\pm$ 3.6 \\
5.14 & $+2.73$ & 29  & 5.0 $\pm$ 2.3  &  13.9 $\pm$  3.4  &  3.8 $\pm$ 1.8  &  7.9 $\pm$  2.8     & 3.9 $\pm$ 2.0  & 15.3 $\pm$ 4.0 \\
5.04 & $-3.37$ & 17  & 3.9 $\pm$ 2.0  &  13.0 $\pm$  3.5  &  2.9 $\pm$ 1.7  &  7.4 $\pm$  2.7     & 3.0 $\pm$ 2.0  & 13.9 $\pm$ 3.5 \\
4.29 & $-3.52$ & 13  & 3.8 $\pm$ 1.8  &  14.0 $\pm$  3.6  &  2.8 $\pm$ 1.7  &  9.5 $\pm$  3.0     & 2.9 $\pm$ 1.5  & 15.2 $\pm$ 3.7 \\
3.98 & $-3.41$ & 20  & 3.9 $\pm$ 2.0  &  15.5 $\pm$  4.1  &  2.9 $\pm$ 1.8  &  11.2 $\pm$  3.4     & 3.0 $\pm$ 1.8  & 16.6 $\pm$ 4.0 \\
3.90 & $-3.19$ & 20  & 4.8 $\pm$ 2.2  &  17.2 $\pm$  4.0  &  3.7 $\pm$ 1.7  &  12.7 $\pm$  3.4     & 3.7 $\pm$ 1.8  & 18.7 $\pm$ 4.1 \\
3.04 & $-3.18$ & 14.5 & 4.9 $\pm$ 2.2 &  19.9 $\pm$  4.4  &  3.8 $\pm$ 1.7  &  17.1 $\pm$  4.1     & 3.8 $\pm$ 1.7  & 21.1 $\pm$ 4.7 \\
2.24 & $-3.72$ & 22  & 3.0 $\pm$ 1.9  &  16.9 $\pm$  4.1  &  2.1 $\pm$ 1.5  &  17.4 $\pm$  4.1     & 2.0 $\pm$ 1.5  & 18.3 $\pm$ 3.9 \\
2.26 & $-3.22$ & 24  & 4.9 $\pm$ 2.3  &  21.7 $\pm$  4.5  &  3.8 $\pm$ 2.0  &  21.8 $\pm$  4.4     & 3.1 $\pm$ 1.9  & 23.5 $\pm$ 4.7 \\
2.15 & $-3.01$ & 32  & 5.7 $\pm$ 2.2  &  24.9 $\pm$  5.0  &  4.0 $\pm$ 1.9  &  24.5 $\pm$  4.8     & 3.9 $\pm$ 1.8  & 26.5 $\pm$ 4.8 \\
2.08 & $-3.59$ & 19  & 3.9 $\pm$ 1.9  &  18.5 $\pm$  4.3  &  2.9 $\pm$ 1.7  &  19.4 $\pm$  4.3     & 2.8 $\pm$ 1.5  & 19.9 $\pm$ 4.5 \\
1.90 & $-2.90$ & 25  & 6.1 $\pm$ 2.5  &  26.7 $\pm$  5.0  &  4.7 $\pm$ 2.1  &  27.9 $\pm$  5.1     & 4.1 $\pm$ 2.0  & 28.7 $\pm$ 5.0 \\
1.77 & $-2.70$ & 33  & 7.1 $\pm$ 2.6  &  30.4 $\pm$  5.4  &  5.2 $\pm$ 2.1  &  31.6 $\pm$  5.4     & 5.0 $\pm$ 2.3  & 32.2 $\pm$ 5.4 \\
1.64 & $-2.52$ & 41  & 8.3 $\pm$ 2.9  &  33.8 $\pm$  5.6  &  6.0 $\pm$ 2.4  &  36.1 $\pm$  6.0     & 5.8 $\pm$ 2.2  & 35.9 $\pm$ 6.0 \\
1.33 & $-2.43$ & 35  & 8.8 $\pm$ 2.8  &  36.6 $\pm$  6.1  &  6.6 $\pm$ 2.5  &  41.0 $\pm$  6.3     & 5.9 $\pm$ 2.5  & 39.1 $\pm$ 5.9 \\
1.08 & $-4.14$ & 18  & 2.2 $\pm$ 1.5  &  15.3 $\pm$  3.8  &  1.9 $\pm$ 1.3  &  17.4 $\pm$  4.0     & 1.9 $\pm$ 1.3  & 16.5 $\pm$ 3.9 \\
0.94 & $-3.69$ & 22  & 3.7 $\pm$ 1.9  &  19.6 $\pm$  4.4  &  2.8 $\pm$ 1.6  &  22.9 $\pm$  4.6     & 2.0 $\pm$ 1.5  & 21.0 $\pm$ 4.2 \\
1.00 & $-3.96$ & 15  & 2.9 $\pm$ 1.7  &  17.0 $\pm$  4.0  &  2.0 $\pm$ 1.4  &  19.6 $\pm$  4.3     & 2.0 $\pm$ 1.3  & 18.1 $\pm$ 4.1 \\
0.88 & $-3.49$ & 25  & 4.0 $\pm$ 1.9  &  21.7 $\pm$  4.6  &  3.0 $\pm$ 1.7  &  25.4 $\pm$  4.8     & 2.9 $\pm$ 1.7  & 23.4 $\pm$ 4.5 \\
0.48 & $-2.26$ & 69  & 10.1 $\pm$ 3.0 &  43.3 $\pm$  6.4  &  7.5 $\pm$ 2.6  &  53.2 $\pm$  7.2     & 6.9 $\pm$ 2.6  & 46.0 $\pm$ 6.7 \\
0.30 & $-2.10$ & 54  & 11.6 $\pm$ 3.1 &  48.1 $\pm$  6.3  &  8.6 $\pm$ 2.8  &  59.6 $\pm$  7.2     & 7.9 $\pm$ 2.7  & 51.0 $\pm$ 6.6 \\
0.21 & $+2.84$ & 47  & 5.0 $\pm$ 2.4  &  26.9 $\pm$  5.0  &  3.9 $\pm$ 2.1  &  31.8 $\pm$  5.2     & 3.7 $\pm$ 1.7  & 28.2 $\pm$ 5.1 \\
-0.01 & $-2.02$ & 70 & 12.0 $\pm$ 3.5 &  50.6 $\pm$  6.4  &  9.0 $\pm$ 2.9  &  63.6 $\pm$  7.0     & 8.5 $\pm$ 2.7  & 53.6 $\pm$ 6.8 \\
-0.09 & $-1.81$ & 68 & 14.2 $\pm$ 3.6 &  57.8 $\pm$  7.4  &  10.7 $\pm$ 3.0  &  73.3 $\pm$  8.1     & 10.1 $\pm$ 3.1  & 61.6 $\pm$ 7.5 \\
-0.09 & $-5.86$ & 6  & 1.0 $\pm$ 0.8  &  6.0 $\pm$  2.4  &  0.9 $\pm$ 0.7  &  7.0 $\pm$  2.5     & 0.0 $\pm$ 0.5  & 6.3 $\pm$ 2.6 \\
-0.31 & $-1.39$ & 90 & 20.7 $\pm$ 4.4 &  77.4 $\pm$  8.5  &  15.2 $\pm$ 3.8  &  98.4 $\pm$  9.7     & 14.8 $\pm$ 3.8  & 81.9 $\pm$ 8.6 \\
-0.28 & $-5.70$ & 8  & 1.0 $\pm$ 0.8  &  6.9 $\pm$  2.6  &  0.9 $\pm$ 0.7  &  7.7 $\pm$  2.8     & 0.0 $\pm$ 0.7  & 7.1 $\pm$ 2.6 \\
-0.34 & $-3.01$ & 46 & 5.8 $\pm$ 2.2  &  29.1 $\pm$  5.4  &  4.2 $\pm$ 2.0  &  35.0 $\pm$  5.5     & 3.9 $\pm$ 1.8  & 30.4 $\pm$ 5.6 \\
-0.48 & $-1.26$ & 79 & 23.3 $\pm$ 4.7 &  83.5 $\pm$  8.8  &  17.4 $\pm$ 4.1  &  105.5 $\pm$  9.9     & 16.7 $\pm$ 4.0  & 88.6 $\pm$ 9.0 \\
-0.63 & $-3.44$ & 30 & 4.0 $\pm$ 1.9  &  22.9 $\pm$  4.5  &  3.0 $\pm$ 1.8  &  27.1 $\pm$  5.0     & 2.9 $\pm$ 1.6  & 24.4 $\pm$ 4.9 \\
-2.41 & $-3.64$ & 33 & 3.1 $\pm$ 1.7  &  18.3 $\pm$  4.2  &  2.1 $\pm$ 1.5  &  17.5 $\pm$  3.9     & 2.1 $\pm$ 1.5  & 19.8 $\pm$ 4.4 \\
-2.56 & $-3.46$ & 29 & 3.9 $\pm$ 2.1  &  19.9 $\pm$  4.1  &  2.9 $\pm$ 1.7  &  17.9 $\pm$  4.1     & 2.9 $\pm$ 1.7  & 21.2 $\pm$ 4.6 \\
-2.86 & $-3.35$ & 24 & 4.1 $\pm$ 2.1  &  20.1 $\pm$  4.2  &  3.0 $\pm$ 1.9  &  17.1 $\pm$  4.2     & 3.0 $\pm$ 1.8  & 21.2 $\pm$ 4.5 \\
-3.16 & $-3.27$ & 34 & 4.3 $\pm$ 2.0  &  19.8 $\pm$  4.3  &  3.1 $\pm$ 1.7  &  16.1 $\pm$  4.1     & 3.0 $\pm$ 1.8  & 21.4 $\pm$ 4.5 \\
-4.99 & $-3.58$ & 19.5  & 3.1 $\pm$ 1.7  &  12.2 $\pm$  3.5  &  2.8 $\pm$ 1.6  &  6.1 $\pm$  2.6     & 2.9 $\pm$ 1.6  & 13.5 $\pm$ 3.5 \\
-6.75 & $-4.55$ & 6.5  & 1.9 $\pm$ 1.2  &  4.9 $\pm$  2.2  &  2.0 $\pm$ 1.0  &  0.9 $\pm$  0.7     & 1.0 $\pm$ 1.0  & 5.2 $\pm$ 2.1 \\
-11.19 & $-2.78$ & 6.0  & 4.9 $\pm$ 2.0  &  0.0 $\pm$  0.0  &  3.9 $\pm$ 1.5  &  0.0 $\pm$  0.0    & 5.0 $\pm$ 2.0  & 0.0 $\pm$ 0.5 \\
\hline

\end{tabular}
\label{tab-MODEL}

\end{table*}

\section{Comparing stellar evolution codes}
\label{AppC}

From P.~Wood's webpage (http://www.mso.anu.edu.au/$\sim$wood/) the
models described in Vassiliadis \& Wood (1993, VW) were downloaded. 
These files list for the individual calculated models on the AGB the
relevant stellar quantities (remaining stellar mass, luminosity and
effective temperature amongst other quantities) and the evolutionary
time. They are available for $Z$ = 0.016 (1.0, 1.5, 2.0, 2.5, 3.5,
5.0 \msol), $Z$ = 0.008 (0.945, 1.0, 1.5, 2.0, 2.5, 3.5, 5.0 \msol), $Z$
= 0.004 (0.89, 1.0, 1.5, 2.0, 2.5, 3.5, 5.0 \msol) and $Z$ = 0.001
(1.0, 1.5 \msol).

For our comparison with simulation we used the solar metallicity models, which 
are expected to be the most representative for our Bulge sample. However, 
different studies indicate that the Bulge may have quite a broad metallicity 
distribution, peaking at about $-0.25$ dex with dispersion of 0.3 dex 
(see e.g. McWilliam \& Rich 1994, Ramirez et al. 2000). 
The AGB lifetime,
 LPV lifetime and
LPV period distribution was determined. The AGB lifetime is defined as
the time between the first model in the file (the start of the AGB)
upto the point where the remaining envelope mass becomes less than
0.04 \msol, or $T_{\rm eff} > 4500 K$, that is taken as the start of
the post-AGB evolution.

For each timestep the fundamental period is calculated following
VW. The star is assumed to be in the Mira instability strip when the
bolometric magnitude is within 0.20 magnitude (the assumed width of
the instability strip at a given period) of the $PL$-relation (Feast
et al. 1989):
\begin{equation}
M_{\rm bol}=  -3.00 \; \log P + 2.85
\end{equation}
assuming a LMC distance of 18.50, {\it and when the mass loss rate is
below a critical value}, as this $PL$-relation is derived for
optically visible objects and the LPV samples studied in G04 and in
the present paper have been culled by only considering objects with
$(J-K)_0 < 2.0$. In such a way the lifetime and the period
distribution of optically visible LPVs can be determined.

The critical mass loss rate is determined by taking for each mass in
the grid of solar metallicities typical values of luminosity and
effective temperature inside the instability strip and then using a
radiative transfer program (Groenewegen 1993) with the appropriate
model atmosphere for M-stars (Fluks et al. 1994), and typical dust
properties (silicate dust, condensation temperature of 1500 K,
dust-to-gas ratio 0.005, expansion velocity 10 \ks) to determine the
critical mass loss rate at which the star would become redder in
$(J-K)$ than 2.0. The critical mass loss rates found are between 
4 and 20 $\times 10^{-6}$ \msolyr\ depending on the initial mass of
the model. In fact, the critical mass loss rate is observed to scale
with $\sqrt{L}$, as expected as the dust optical depth predominantly
determines the infrared colours, hence the mass loss rate is
proportional to the stellar radius, all other things being equal. For
the 1.0 and 1.5\msol\ initial mass models the mass loss rate inside
the instability strip always remains below the critical mass loss rate. 
The adopted critical mass loss rate is 1.0 $\times 10^{-5}\, \sqrt{L/13000}$ \msolyr.
The results for the solar metallicty models are summarised in Table~\ref{tab-VW}

We will now consider the synthetic AGB models of Wagenhuber \& Groenewegen (1998; WG). 
The reason being that the VW models exist only for a limited number of
initial masses and secondly because mass loss on the RGB was only
included for initial masses {\it below} 1\msol, while it is well known
that the effect of mass loss on the RGB is substantial also at and
above 1 \msol. The VW mass loss rate recipe was implemented, and to
mimic the VW tracks as closely as possible the mixing length parameter
$\alpha$ (basically setting the effective temperature scale) in the WG
models was tuned to give similar AGB and LPV lifetimes.  The results
are summarised in Table~\ref{tab-VW}. It turns out that with $\alpha = 2.0$ 
the lifetimes match very well especially at low initial mass.

\begin{figure*}
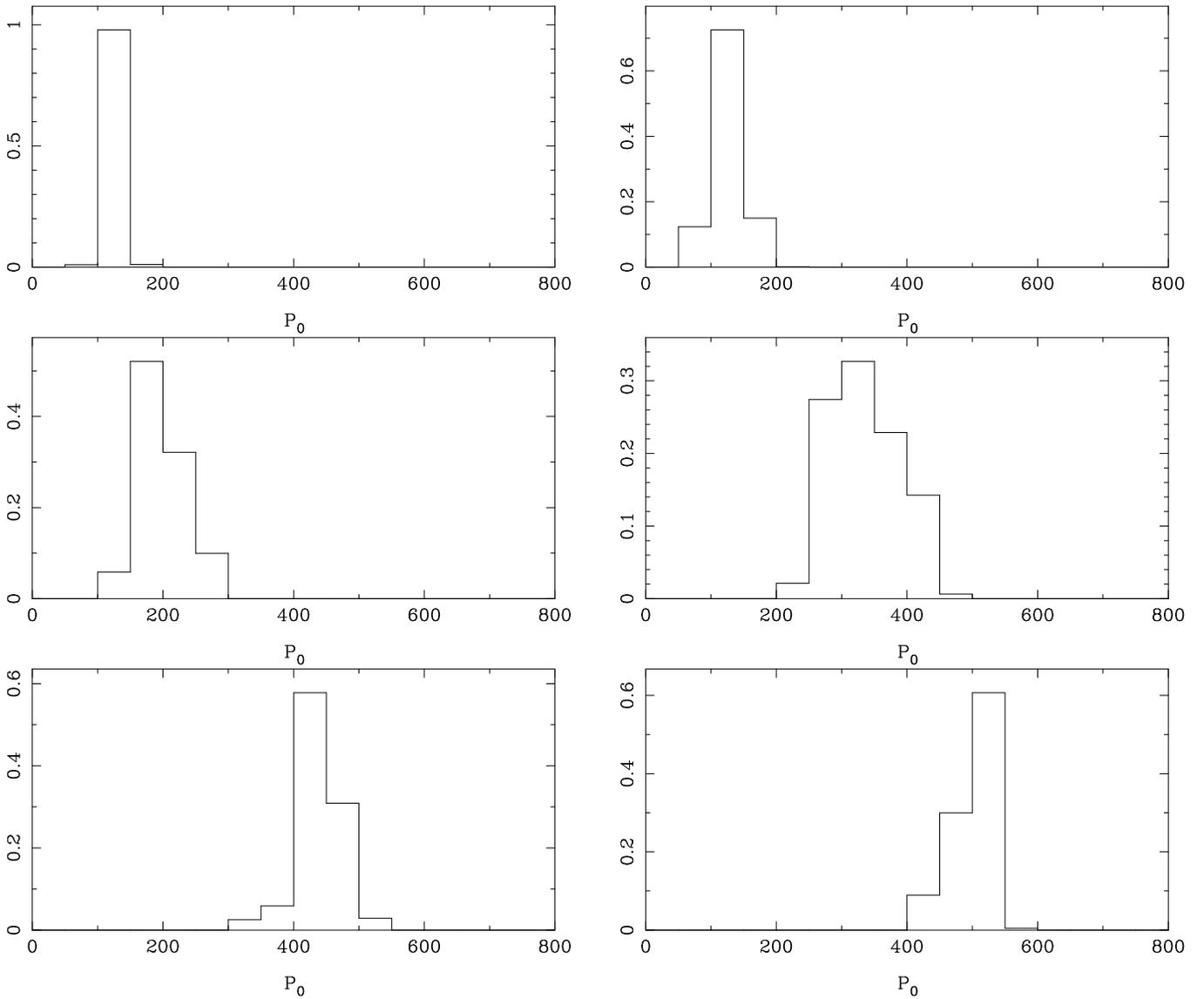


\begin{minipage}{0.48\textwidth}
\resizebox{\hsize}{!}{\includegraphics{Pdistr_1.1.ps}}
\end{minipage}
\hfill
\begin{minipage}{0.48\textwidth}
\resizebox{\hsize}{!}{\includegraphics{Pdistr_1.2.ps}}
\end{minipage}

\begin{minipage}{0.48\textwidth}
\resizebox{\hsize}{!}{\includegraphics{Pdistr_1.5.ps}}
\end{minipage}
\hfill
\begin{minipage}{0.48\textwidth}
\resizebox{\hsize}{!}{\includegraphics{Pdistr_2.0.ps}}
\end{minipage}

\begin{minipage}{0.48\textwidth}
\resizebox{\hsize}{!}{\includegraphics{Pdistr_2.5.ps}}
\end{minipage}
\hfill
\begin{minipage}{0.48\textwidth}
\resizebox{\hsize}{!}{\includegraphics{Pdistr_3.0.ps}}
\end{minipage}

\caption[]{
Theoretical period distribution of optically visible stars inside the observed
instability strip for masses 1.1, 1.2, 1.5, 2.0, 2.5, 3.0\msol\ (left to
right, top to bottom).
}
\label{Fig-PerDistr}
\end{figure*}

Mass loss on the RGB is described by a Reimers law with a scaling
factor $\eta_{\rm RGB} = 0.35$. This gives the required mass loss
(0.13, 0.16, 0.17 \msol\ for a star of 1.0\msol\ initial mass at $Z =$
0.004, 0.008, 0.019, respectively; M.~Salaris, private communication)
to give the observed mean colour of Horizontal Branch stars in
Galactic Globular Clusters. Table~\ref{tab-WG} summarises for a set of
WG models with $\eta_{\rm RGB} = 0.35$ and $\alpha = 2.0$ the AGB and
LPV lifetime, and Figure~\ref{Fig-PerDistr} displays the period
distribution of optically visible stars inside the instability strip
(normalised to one each time) for a few initial masses.  

\begin{figure*}
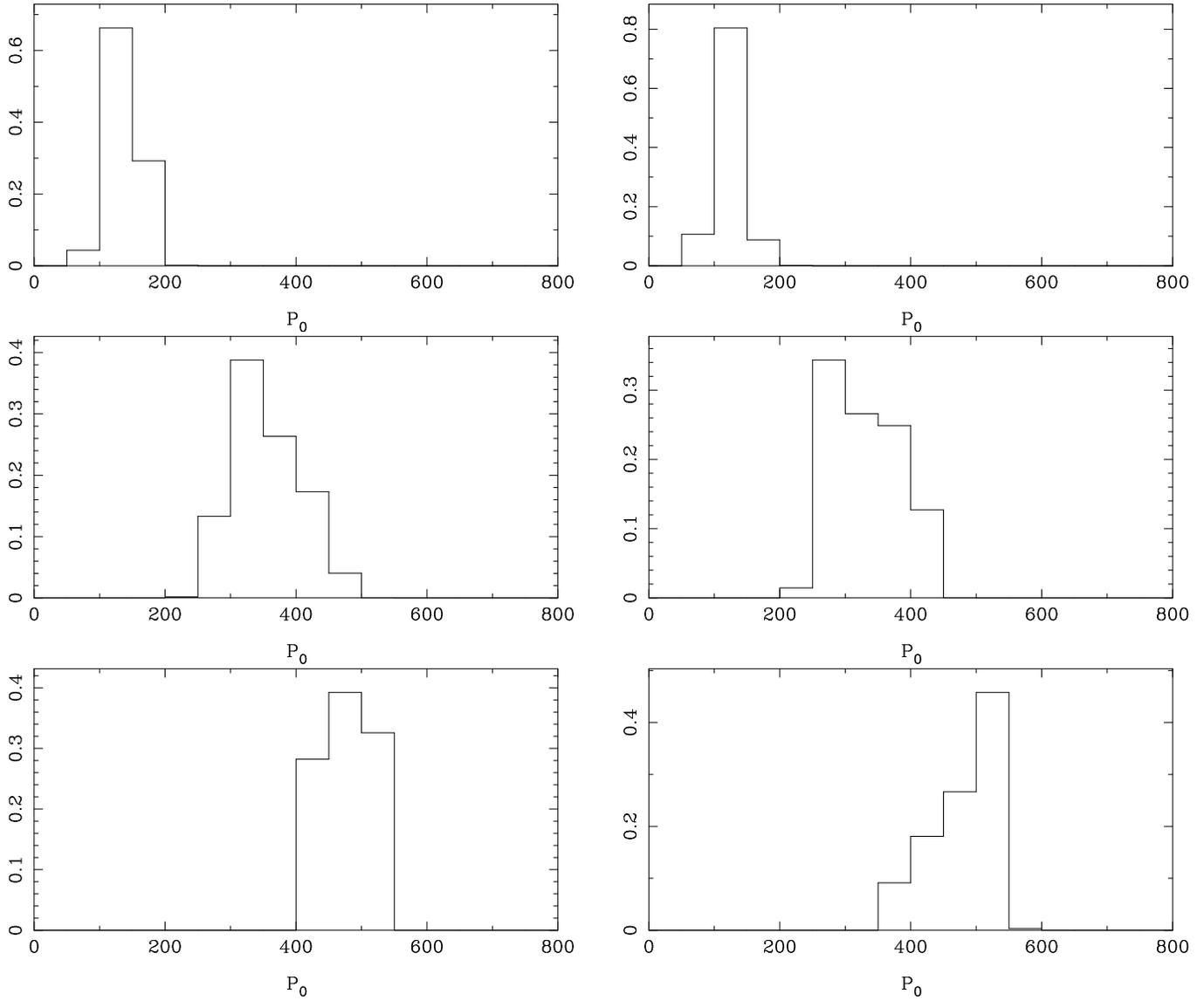


\begin{minipage}{0.48\textwidth}
\resizebox{\hsize}{!}{\includegraphics{Pdistr_1.2_Z=0.010.ps}}
\end{minipage}
\hfill
\begin{minipage}{0.48\textwidth}
\resizebox{\hsize}{!}{\includegraphics{Pdistr_1.2_Z=0.020.ps}}
\end{minipage}

\begin{minipage}{0.48\textwidth}
\resizebox{\hsize}{!}{\includegraphics{Pdistr_2.0_Z=0.010.ps}}
\end{minipage}
\hfill
\begin{minipage}{0.48\textwidth}
\resizebox{\hsize}{!}{\includegraphics{Pdistr_2.0_Z=0.020.ps}}
\end{minipage}

\begin{minipage}{0.48\textwidth}
\resizebox{\hsize}{!}{\includegraphics{Pdistr_3.0_Z=0.010.ps}}
\end{minipage}
\hfill
\begin{minipage}{0.48\textwidth}
\resizebox{\hsize}{!}{\includegraphics{Pdistr_3.0_Z=0.020.ps}}
\end{minipage}

\caption[]{
As Figure~\ref{Fig-PerDistr} illustrating the influence of
metallicity on the theoretical distribution. 
In the left hand panels for $Z = 0.010$ and on the right
for $Z = 0.020$, for stars with inital masses 1.2 (top), 2.0 (middle)
and 3.0 (bottom) \msol.
}
\label{Fig-PerDistr-Z}
\end{figure*}

Table~\ref{tab-WG} also includes for a few initial masses the
results if slightly different metallicities of $Z$= 0.01 and 0.02 are
adopted, and Figure~\ref{Fig-PerDistr-Z} shows the corresponding
Period distribution. These results indicate that the effect of
metallicity on the pulsation properties for the typical metallicities
in the Bulge is small.  \\

\begin{table*}
\caption{AGB lifetime and optically visible LPV lifetimes for two sets of models}
\begin{tabular}{cccccccc} \hline
        &         & \multicolumn{3}{c}{Vassiliadis \& Wood} & 
                                                      \multicolumn{3}{c}{Wagenhuber \& Groenewegen}  \\ 
 $Z$    & Mass    & MS lifetime & AGB lifetime   & LPV lifetime  &   AGB lifetime   & LPV lifetime & remark  \\
        & (\msol) & (10$^9$ years)  & (10$^3$ years) & (10$^3$ years) & (10$^3$ years)  & (10$^3$ years) & \\ \hline
0.016 & 1.0 &  11.25 & 595 & 101  &  487 &  49 & $\eta_{\rm RGB}= 0$, $\alpha$= 1.9 \\
      &     &      &      &       & 560 &  93 & $\eta_{\rm RGB}= 0$, $\alpha$= 2.0 \\
      &     &      &      &       & 595 & 129 & $\eta_{\rm RGB}= 0$, $\alpha$= 2.1 \\
0.016 & 1.5 & 2.742 & 929 & 272  &  873 & 303 & $\eta_{\rm RGB}= 0$, $\alpha$= 1.9 \\
      &     &      &      &       & 942 & 284 & $\eta_{\rm RGB}= 0$, $\alpha$= 2.0 \\
      &     &      &      &       & 1019 & 282 & $\eta_{\rm RGB}= 0$, $\alpha$= 2.1 \\
0.016 & 2.0 & 1.236 & 1168 & 153  &  1596 & 253 & $\eta_{\rm RGB}= 0$, $\alpha$= 2.0 \\
      &     &      &      &       & 1684 & 222 & $\eta_{\rm RGB}= 0$, $\alpha$= 2.1 \\
0.016 & 2.5 &  0.619 & 2291 &   8  & 2284 &  84 & $\eta_{\rm RGB}= 0$, $\alpha$= 2.0 \\ 
      &     &      &      &       & 2354 &  50 & $\eta_{\rm RGB}= 0$, $\alpha$= 2.1 \\
0.016 & 3.5 & 0.231 &  543 &   0  &  497 &   0 & $\eta_{\rm RGB}= 0$, $\alpha$= 2.0 \\ 
      &     &      &      &       & 542 &   0 & $\eta_{\rm RGB}= 0$, $\alpha$= 2.1 \\
0.016 & 5.0 &  0.096 & 673 &   0  & 175 &   0 & $\eta_{\rm RGB}= 0$, $\alpha$= 2.0 \\ 
      &     &      &      &       & 174 &   0 & $\eta_{\rm RGB}= 0$, $\alpha$= 2.1 \\
\hline
\end{tabular}
\label{tab-VW}
\end{table*}

\begin{table}
\caption{AGB lifetime and optically visible LPV lifetimes for the
final set of synthetic models, with $\eta_{\rm RGB} = 0.35$ and $\alpha = 2.0$.}
\begin{tabular}{rrrr} \hline
 $Z$    & Mass    &  AGB lifetime   & LPV lifetime    \\
        & (\msol) & (10$^3$ years)  & (10$^3$ years)  \\ \hline
0.016 & 0.85 &  240 &  1.6 \\
0.016 & 0.9 &  288 &   1.0 \\
0.016 & 1.0 &  305 &   1.5 \\
0.016 & 1.1 &  445 &  41.  \\
0.010 & 1.2 &  644 & 132.  \\
0.016 & 1.2 &  552 & 108.  \\
0.020 & 1.2 &  528 &  95.   \\
0.016 & 1.3 &  648 & 161.  \\
0.016 & 1.4 &  770 & 247.  \\
0.016 & 1.5 &  858 & 296.  \\
0.016 & 1.6 &  965 & 293.  \\
0.016 & 1.7 & 1103 & 282.  \\
0.016 & 1.8 & 1211 & 296.  \\
0.016 & 1.9 & 1362 & 289.  \\
0.010 & 2.0 & 1610 & 234. \\
0.016 & 2.0 & 1522 & 269.  \\
0.020 & 2.0 & 1527 & 283.  \\
0.016 & 2.1 & 1711 & 250.  \\
0.016 & 2.2 & 1898 & 220.  \\
0.016 & 2.3 & 2071 & 182.  \\
0.016 & 2.4 & 2182 & 131.  \\
0.016 & 2.5 & 2269 &  90.  \\
0.016 & 2.6 & 2300 &  59.  \\
0.016 & 2.7 & 2260 &  40.  \\
0.016 & 2.8 & 2125 &  26.  \\
0.016 & 2.9 & 1903 &  14.  \\
0.010 & 3.0 & 1435 &   3.  \\
0.016 & 3.0 & 1643 &  10. \\
0.020 & 3.0 & 1748 &  10. \\
0.016 & 3.1 & 1355 &   3. \\
0.016 & 3.4 &  641 &   0. \\
\hline
\end{tabular}
\label{tab-WG}
\end{table}

\end{document}